\documentclass[onecolumn]{article}
\pdfoutput=1
                        \newif\ifpaper \newif\ifPDF               
                        \newif\ifOUP \newif\ifboyscout            
                        \newif\ifdasbuch \newif\ifarticle         
                        \newif\ifsolutions                        
                        \dasbuchtrue 
                        \boyscouttrue 
                        \solutionstrue 
                        \paperfalse\PDFtrue 
                        \OUPfalse \articlefalse  

\usepackage{nolta2013}
\usepackage{txfonts}
\usepackage[latin1]{inputenc}
\usepackage[T1]{fontenc}
\usepackage{times}
\usepackage[pdftex]{graphicx}
\usepackage{array}
\usepackage[pdftex,colorlinks]{hyperref}
\usepackage{datetime}
\usepackage{lineno}

\newcommand*{\action}{{\mathcal{X}}}
\newcommand{\beq}{\begin{equation}}
\newcommand{\continue}{\nonumber \\ }

\newcommand{\eeq}{\end{equation}}
\newcommand{\ee}[1] {\label{#1} \end{equation}}
\newcommand{\bea}{\begin{eqnarray}}

\newcommand{\eea}{\end{eqnarray}}
\newcommand{\barr}{\begin{array}}
\newcommand{\earr}{\end{array}}
\newcommand{\msr}{\ensuremath{\rho}} 
\newcommand{\transp}[1]{{#1}{}^\top}
\newcommand{\orbitDist}{\ensuremath{z}}
\newcommand{\diffTen}{\ensuremath{\Delta}}  
\newcommand{\covMat}{\ensuremath{Q}}             
\newcommand{\refeq}  [1] {(\ref{#1})}
\newcommand{\rf}     [1] {~\cite{#1}}
\newcommand{\Lyap}{\ensuremath{\lambda}} 
\newcommand{\Fokker}{Fokker-Planck}
\newcommand{\eigenvL}{\ensuremath{s}}      

\newcommand{\xp}{\mathbf{x}}

\begin{document}

\title{Towards the Resolution of a Quantized Chaotic Phase Space: The Interplay of Dynamics with Noise}

\author{Domenico Lippolis$^{1}$ and Akira Shudo$^{2}$}

\address{ 
$^{1}$ \quad Institute for Applied Systems Analysis, Jiangsu University, Zhenjiang 212013, China; domenico@ujs.edu.cn\\
$^{2}$ \quad Department of Physics, Tokyo Metropolitan University, Minami-Osawa, Hachioji, Tokyo 192-0397, Japan}

\maketitle

\abstract{We outline formal and physical similarities between the quantum dynamics of open systems, and the mesoscopic description of 
classical systems affected by weak noise. The main tool of our interest is the dissipative Wigner equation, that, for suitable timescales, becomes
analogous to the Fokker-Planck equation describing classical advection and diffusion. This correspondence allows in principle to surmise
a finite resolution, other than the Planck scale, for the quantized state space of the open system, particularly meaningful when the latter underlies
chaotic classical dynamics. We provide representative examples of the quantum-stochastic parallel with noisy Hopf cycles and Van der Pol type oscillators.}

\section{Introduction}

Efforts to reconcile classical and quantum mechanics are just about as old as quantum mechanics itself. While the formulation in Hilbert space makes it difficult to establish a direct correspondence
between the two, a projection of the wave function to phase space may reveal some formal affinities between the quantum evolution of probability density and the traditional Liouville 
formalism of classical mechanics.
The closest one can get to relate the two is by projecting the Liouville-von Neumann equation onto a suitable state space. For example, choosing the traditional 
phase space, we may obtain the so-called Wigner representation, that shares similarities with the aforementioned classical density evolution.

Yet, there are also notable differences, as it stands to reason. The Wigner function, that is the projection of the density operator onto the phase-space, may also take on
negative values, its evolution is governed by an equation plagued with an infinity of derivatives, and, as an indirect consequence of that,
it may attain scales smaller than Planck's constant~\cite{Zurek01}. This is especially true in systems whose underlying classical dynamics exhibits chaotic behavior.

In reality, however, no system is perfectly and eternally isolated, and exchange of matter or energy with the surrounding environment is inevitable,
whether due to measurements, thermal interactions, or shot noise~\cite{BreuPetr,ZubScul,Gard}.
That brings dissipation into the picture, and with that, decoherence.

The effect of the environment on the evolution of a density matrix in a phase-space representation was first 
studied by Feynman and Vernon~\cite{FeyVer63}, who extended the path-integral formalism to dissipative quantum dynamics.
Later, Caldeira and Leggett~\cite{CalLeg83} derived an equivalent partial differential equation for the density matrix, that bears diffusive and
dissipative terms, similarly to a Fokker-Planck equation. The latter analogy was then thought to hold but in the semiclassical limit,
until a new wave of contributions~\cite{DitGra88,ZurPaz94,Kolov94,Kolov96} reexamined the problem in a quantum chaotic setting. A most remarkable outcome of those 
works is the identification of a decoherence time, beyond which the Wigner equation is in all a Fokker-Planck equation since     
the higher-order derivatives may be safely neglected, and the quantized phase space may be resolved only up to a finite scale. 
Such resolution does not depend on the Planck constant, but rather emerges from the balance of the phase-space contraction
rate (Lyapunov exponent) with the coupling of the system with an Ohmic environment.

More recent contributions have focussed on the efficiency of Wigner evolution 
for general types of dissipation~\cite{Cabrera15}, and on
obtaining a Lindblad-based dissipative Wigner equation to tackle  
quantum friction~\cite{Bondar16,CarloQFric19}.

Once it is established that, under suitable conditions and after a sufficiently long time of evolution, the Wigner equation has the 
form of a Fokker-Planck equation, the quantum dissipative problem is cast into a classical stochastic 
process. Moreover, if the underlying classical dynamics of the quantum system in exam is chaotic,  
the limiting resolution of the phase space postulated in refs.~\cite{ZurPaz94,Kolov96} is
not expected to be uniform, but it will depend on the local interplay of the stretching/contraction     
with the dissipation. In the equivalent classical noisy problem, it is the `Brownian' diffusion that plays 
the role of the dissipation.

In the past decade, significant steps~\cite{LipCvi10,CviLip12,noltino,HenLipCvi15,HenLipCvi18} were taken to determine the resolution
of a chaotic state space in the presence of weak noise, and so reduce the dynamics
to a Markov process of finite degrees of freedom, in the form of a connectivity matrix. 
Low-dimensional discrete-time dynamical systems such as logistic- or H\'enon-type maps have been treated
in a non-Hamiltonian setting, whose quantum analogs are in principle difficult to identify.
The optimal resolution hypothesis should be extended to continuous-time flows as well,
and the starting point of that roadmap is a thorough comprehension of the
steady-state solutions of the Fokker-Planck equation around the building blocks of chaos: periodic orbits.

Here, we intend to lay the foundation of that understanding, by solving the Fokker-Planck
equation of nonlinear paradigmatic dynamical systems, classical and with weak noise.
We examine two-dimensional flows featuring nonlinearities but not yet chaos, where the competition 
between contraction and noise around a limit cycle results in a stationary density, which characterizes
the steady state, and, as shown at the very end of the manuscript, shares common traits with the steady-state Wigner function of 
a case study in quantum dissipative dynamics.

The article is structured as follows: in section~\ref{Qconcepts} we review the basic tools of the 
phase-space representation of quantum dynamics, both in closed and open systems.
We follow up in section~\ref{Qchaos}  by discussing the main issues related to the evolution of the Wigner function in a 
quantum chaotic setting, the effects of dissipation, and the correspondence of the Wigner- with the Fokker-Planck equation.
In section~\ref{ContrDiff} a novel methodology is introduced 
to evaluate the steady-state solution of the Fokker-Planck equation around a periodic orbit, which casts the partial differential equation
into an ordinary differential equation for the covariance matrix, known as Lyapunov equation.  
We first present a proof of concept on the simplest limit cycle,
of circular shape as from a Hopf bifurcation, to be followed in section~\ref{VdP} by the treatment of the nonlinear oscillators
that are the main object of the current study. At the end of the section the results on the Fokker-Planck steady-state densities
are paralleled to those obtained for the Wigner function in a recent study of a quantum-dissipative model of the same oscillators.
Summary and discussion close the paper.  

\section{Density matrix, Wigner function, and dissipation}
\label{Qconcepts}

Given a collection of physical systems, the ensemble average of an observable $A$ is given by
\begin{linenomath}
\begin{equation}
\left< A \right> = \sum_i \rho_i \langle \psi_i| A| \psi_i\rangle
\,,
\end{equation}
\end{linenomath}
or, using
\begin{linenomath}
\begin{equation}
\rho = \sum_i \rho_i |\psi_i\rangle \langle \psi_i|
\,,
\end{equation}
\end{linenomath}
one can simply write
\begin{linenomath}
\begin{equation}
\left< A \right> = \mathrm{Tr} \left[\rho\,A\right]
\,,
\end{equation}
\end{linenomath}
so that, if the observable $A$ is time-independent, knowing $\rho$ at all times means solving the problem of 
dynamics. That is the motivation for studying the density matrix $\rho$ in the first place.

\subsection{Quantum dynamics in the phase space}

Now, the density $\rho$ evolves according to the Liouville-von Neumann equation
\begin{linenomath}
\begin{equation}
i\hbar \rho_t = \left[\rho, H\right]
\,,
\label{LvN}
\end{equation}
\end{linenomath}
the quantum analog of the well-known Liouville equation
\begin{linenomath}
\begin{equation}
\rho_t = \left\{\rho, H\right\}
\end{equation}
\end{linenomath}
of classical dynamics, which we spell out in phase space:
\begin{linenomath}
\begin{equation}
\partial_t\rho = -\frac{p}{m}\partial_x\rho + \left(\partial_x V(x)\partial_p\right) \rho 
\,,
\label{LiouvEq}
\end{equation}
\end{linenomath}
assuming a Hamiltonian of the form $H= \frac{p^2}{2m} + V(x)$.

In order to integrate the Liouville-von Neumann equation~(\ref{LvN}), we need to project it onto some basis,
and several representations are already available to us, for instance, the P- or the 
Q-representation (a.k.a. Husimi's)
 \begin{linenomath}
\begin{equation}
Q(\alpha,\alpha^*) = \frac{1}{\pi}\langle \alpha|\rho|\alpha\rangle
\,,
\end{equation}
\end{linenomath}
with $|\alpha\rangle$ a coherent state. Studying quantum-to-classical correspondence, especially of a system that
exhibits chaotic behavior, is generally best achieved by using the Wigner representation~\cite{Case08}
 \begin{linenomath}
\begin{equation}
W(x,p) = \frac{1}{2\pi\hbar}\int e^{-ipy/\hbar} \psi\left(x+\frac{y}{2}\right)\psi^*\left(x-\frac{y}{2}\right)dy
\,,
\end{equation}
\end{linenomath}
which can also be expressed in terms of the density matrix, as
 \begin{linenomath}
\begin{equation}
W(x,p) = \frac{1}{2\pi\hbar}\int e^{-ipy/\hbar} \langle x+y/2|\rho|x-y/2\rangle dy
\,,
\end{equation}
\end{linenomath}
the operation being called Weyl transform. An operator $A$ may also be projected onto 
phase space, by applying a Weyl transform:
 \begin{linenomath}
\begin{equation}
\tilde{A}(x,p) =  \frac{1}{2\pi\hbar}\int e^{-ipy/\hbar} \langle x+y/2|A|x-y/2\rangle dy
\,,
\end{equation}
\end{linenomath}
which can prove handy in the evaluation of expectation values, that is
\begin{linenomath}
\begin{equation}
\left< A \right> = \mathrm{Tr} \left[\rho\,A\right] = \int W(x,p) \tilde{A}(x,p) dxdp
\,,
\end{equation}
\end{linenomath}
since, in general,
\begin{linenomath}
\begin{equation}
\mathrm{Tr} \left[A B\right] = \int \tilde{A}(x,p) \tilde{B}(x,p) dxdp
\,.
\end{equation}
\end{linenomath}
Thus, expectation values of observables are determined by means of phase-space averages, and the problem of
quantum mechanics boils down to that of the time evolution of the Wigner function. It has been shown~\cite{Case08} that 
$W(x,p)$ obeys the Wigner equation
\begin{linenomath}
\begin{equation}
\partial_t W(x,p) = -\frac{p}{m}\partial_x W(x,p) + \sum_{s=0}^\infty c_s (-\hbar^2)^s \partial_x^{2s+1} V(x)\partial_p^{2s+1} W(x,p)
\,,
\label{WigEq}
\end{equation}
\end{linenomath}
(with $c_s=\frac{2^{-2s}}{(2s+1)!}$)  
that, in general, bears an infinite number of terms. In reality, integrating equation~(\ref{WigEq}) can already be impractical
if there are just a few nontrivial terms in the summation~\cite{AltHaa12}. If the potential $V(x)$ is at most quadratic, the Wigner
equation reduces to Liouville's, as in~(\ref{LiouvEq}). Otherwise, Eq.~(\ref{WigEq}) is still not easy to deal with, and, importantly, 
it may not be truncated in the semiclassical limit, since the terms $\partial_p^{2s+1} W(x,p)$ bring down powers of $\hbar^{-1-2s}$,
so that
\begin{linenomath}
\begin{equation}
\nonumber
\hbar^{2s} \cdot \frac{1}{\hbar^{2s+1}} \sim \hbar^{-1} 
\,,
\end{equation}
\end{linenomath}
and $O\left(\hbar^{-1}\right)$ does grow in the limit $\hbar\rightarrow0$, making no terms in the Wigner equation negligible, in
principle.

\subsection{Open systems}
\label{OpSyst}

On the other hand, let us suppose the system is connected to an environment, whose interaction produces two additional
terms in the right-hand side of the Wigner equation~(\ref{WigEq}), that is~\cite{ZurPaz94}
\begin{linenomath}
\begin{equation}
\nonumber
2\gamma\partial_p \left[pW(x,p)\right] + D\partial_{pp}^2 W(x,p)
\,.
\label{DissTerms}
\end{equation}
\end{linenomath} 
The first term produces relaxation, due to the exchange of energy with the environment, and $\gamma$ is the 
relaxation rate. The second term means diffusion, responsible for the so-called decoherence process, where one 
sets $D=2\gamma Mk_BT$, with $M$ mass of the system, and $T$ temperature of the environment. 
The dissipation and diffusion terms are obtained from a
path-integral formulation of the system-environment interaction, that traces back to the works of 
Feynman and Vernon~\cite{FeyVer63}, and, later, of Caldeira and Leggett~\cite{CalLeg83}.  

If the potential $V(x)$ is at most quadratic, one recovers the Fokker-Planck equation, that describes the classical
evolution of the density of trajectories produced by a particle subject to Brownian motion:
\begin{linenomath}
\begin{equation}
\partial_t W(x,p) = -\frac{p}{m}\partial_x W(x,p) + \partial_x V(x)\partial_p W(x,p) + 2\gamma\partial_p \left[pW(x,p)\right] + D\partial_{pp}^2 W(x,p)
\,.
\label{DissWigEq}
\end{equation}
\end{linenomath}  
This equation is fully quantum mechanical, and $W(x,p)$ may take on negative values, unlike the classical phase-space density of 
a Brownian particle.

Yet, for a general potential $V(x)$, the evolution of the dissipative system is ruled by the full-fledged Wigner equation~(\ref{WigEq}) plus
the terms~(\ref{DissTerms}) due to the environment: 
\begin{equation}
\partial_t W(x,p) = -\frac{p}{m}\partial_x W(x,p) + \sum_{s=0}^\infty c_s (-\hbar^2)^s \partial_x^{2s+1} V(x)\partial_p^{2s+1} W(x,p) 
+ 2\gamma\partial_p \left[pW(x,p)\right] + D\partial_{pp}^2 W(x,p)
\,.
\label{WigEqDiss}
\end{equation}
The resulting equation is still plagued with an infinite number of derivatives, and is thus of impractical integration. In the next section, we 
discuss whether and how it is safe to neglect the higher-order terms in Eq.~(\ref{WigEqDiss}), in the context of quantum chaos.

\section{Stretching, contracting, and Zaslavsky's time}
\label{Qchaos}

Let us examine some aspects of the evolution of the Wigner function, when the underlying classical dynamics of the
system is chaotic. By established knowledge~\cite{GaspBook}, the two main features of chaos are
\begin{enumerate}
\item	Nearby trajectories diverge exponentially fast, meaning that, letting $\mathbf{x}=(x,p)$,
\begin{linenomath}
\begin{equation}
\lambda = \lim_{t\rightarrow\infty} \ln \left| \frac{\delta \mathbf{x}(t)}{\delta \mathbf{x}(0)}\right| > 0
\,,
\end{equation}
\end{linenomath}  
in other words, the difference $\delta \mathbf{x}(t)$ between any two nearby trajectories grows exponentially fast for any initial condition.
This feature is also described as extreme sensitivity of the system to initial conditions. 
\item The number $M$ of qualitatively distinct orbits (`configurations', tagged by symbolic sequences) scales exponentially with their length, so that the topological entropy is positive:
\begin{linenomath}
\begin{equation}
S = \lim_{t\rightarrow\infty} \frac{1}{t}\ln M(t) > 0
\,.
\end{equation}
\end{linenomath}    
\end{enumerate}

\subsection{Chaos and the Wigner function}

Chaos is the result of a stretching and folding process mainly due to nonlinearities. For a Hamiltonian system,
volumes in the phase space are conserved (by Liouville's theorem), so that the amount of stretching
(diverging trajectories) in some directions must be compensated by an equal amount of contraction in others. 

As a result, the inconvenient higher-order terms in the Wigner equation~(\ref{WigEq}) can be estimated to 
evolve as
\begin{linenomath}
\begin{equation}
\partial_p^{2s+1} W(x,p) \propto \frac{W(x,p)}{\delta p^{2s+1}(t)} \sim
 \frac{W(x,p)}{\delta p(0)\,e^{-(2s+1)\lambda t}}
\,,
\end{equation}
\end{linenomath}  
for smooth enough $W(x,p)$. Thus, the inherent problem is in principle not with the smoothness of the density,
but rather with the fact that the phase space contracts at an exponential rate, and therefore the contribution 
of higher-order derivatives in the equation~(\ref{WigEq}) is more and more important, as time proceeds.  
To better illustrate that, let us compare the terms in the Poisson brackets of the Liouville equation~(\ref{LiouvEq})
(also present in the full-fledged Wigner equation), with the higher-order terms in Eq.~(\ref{WigEq}):
\begin{linenomath}
\begin{equation}
\frac{\partial_x V(x)\partial_p W(x,p)}{c_s\partial_x^{2s+1}V(x)\partial_p^{2s+1}W(x,p)} 
\sim \frac{1}{c_s}\frac{\partial_x V(x)}{\partial_x^{2s+1}V(x)}\delta p^{2s}
\sim \frac{1}{c_s}\frac{\partial_x V(x)}{\partial_x^{2s+1}V(x)}\delta p(0)\,e^{-2s\lambda t} \gg1 
\,
\label{WigSimpCond}
\end{equation}
\end{linenomath}  
is a condition for the higher-order terms to be negligible with respect to the lower-order, `Liouville' terms.
The above inequality can be inverted, and turned into a condition for the time $t$: 
\begin{linenomath}
\begin{equation}
t \ll  \frac{1}{\lambda}\ln\left[\frac{\partial_x V(x)}{\partial_x^{2s+1}V(x)}\frac{\delta p(0)}{c_s}\right]^{-1/2s}
\,.
\end{equation}
\end{linenomath}  
Identifying the quantity $\action_V\,\delta p(0)=\frac{\partial_x V(x)\delta p(0)}{\partial_x^{2s+1}V(x)}$ with the typical 
action of the system, we can now understand 
\begin{linenomath}
\begin{equation}
t^* \sim \frac{1}{\lambda}\ln\frac{\action_V\,\delta p(0)}{\hbar}
\label{Zast}
\end{equation}
\end{linenomath}  
as the time scale within which the inconvenient higher-order terms ($s\geq1$) of the Wigner equation~(\ref{WigEq}) may be neglected.     
Some literature refers to $t^*$ as Zaslavsky's time~\cite{BerZas78}. Its meaning is somehow related to the more commonly
mentioned Ehrenfest time, as in fact $t^*$ is longer the larger the ratio of the typical action to $\hbar$,
and longest in the semiclassical limit. Still, the basic idea of this correspondence time does not relate directly with interference or need 
`semiclassical' dynamics, but rather implies a finite resolution for the quantized phase space within a certain time scale, irrespective of 
the scale of the action. In general,
the smoother the potential $V(x)$, the longer $t^*$, whereas the larger the Lyapunov exponent $\lambda$, the 
shorter $t^*$.

\subsection{A resolution for the quantized phase space}  
In a simplified but physically meaningful description, that will then prove more accurate
as a local model, we may recognize and estimate the competing effects of dynamical contraction on the one hand, and 
of dissipation-induced diffusion on the other hand, by quantizing the Hamiltonian $H=\lambda xp$. A 
wave packet of the form 
\begin{linenomath}
\begin{equation}
W(x,p) \sim e^{-x^2/\sigma^2 -\sigma^2 p^2}
\end{equation}
\end{linenomath}
evolves separately along the stretching $x-$direction, and the contracting $p-$direction.
In momentum space, we have the Schr\"odinger equation
\begin{linenomath}
\begin{equation}
\partial_t u(p,t) = \lambda p\partial_{p} u(p,t) \Rightarrow 
u(p,t) = u_0\left(pe^{\lambda t}\right)
\,,
\end{equation}
\end{linenomath}    
that maps the wave packet in the $p-$direction as
\begin{linenomath}
\begin{equation}
e^{-\sigma^2 p^2/2} \rightarrow 
e^{e^{2\lambda t}\sigma^2 p^2/2}
\,,
\end{equation}
\end{linenomath}    
and thus the width $\sigma^{-2}$ shrinks by a factor of $e^{-2\lambda t}$ after time $t$. 
Identifying $\sigma^{-1}$ with the uncertainty $\delta p(t)$ of the momentum, we may say that  
\begin{linenomath}
\begin{equation}
\delta p(t) \sim \delta p(0)e^{-\lambda t}
\end{equation}
\end{linenomath} 
along the contracting direction.
On the other hand, connecting the system to an environment brings about diffusion, and
\begin{linenomath}
\begin{equation}
\partial_t u(p,t) = D\partial_{pp} u(p,t) \Rightarrow 
u(p,t) \sim e^{-p^2/2(\delta p(0) + Dt)}
\,,
\end{equation}
\end{linenomath} 
whose variance evolves as $\sqrt{Dt}$:
\begin{linenomath}
\begin{equation}
\delta p(t) \sim \left[\delta p(0) + Dt\right]^{1/2}
\label{DiffEvol}
\end{equation}
\end{linenomath}  
Then, intuitively, there must be some minimal scale in the contracting direction,
set by
\begin{linenomath}
\begin{equation}
\delta p_{\mathrm{min}} \sim \left[\frac{D}{2\lambda}\right]^{1/2}
\,.
\end{equation}
\end{linenomath} 
 The full picture is called Ornstein-Uhlenbeck problem~\cite{Risken96}
\begin{linenomath}
\begin{equation}
\partial_t u(p,t) = D\partial_{pp} u(p,t) - \lambda\partial_p u(p,t)
\,.
\end{equation}
\end{linenomath} 
In particular, the larger $\delta p_{\mathrm{min}}$, the closer the evolution of 
the Wigner function to a stochastic process. More precisely, the regime  
where we may neglect the higher-order derivatives in the Wigner equation is deduced from
Eq.~(\ref{WigSimpCond}) as
\begin{linenomath}
\begin{equation}
\frac{\action_V\, \delta p_{\mathrm{min}}}{\hbar} \gg 1
\,,
\end{equation}
\end{linenomath}  
and that requires a  relatively smooth potential, and the coefficient of the decoherence term in Eq.~(\ref{DissWigEq}), $D$,
to be comparable to the Lyapunov exponent $\lambda$.  
Now, if $\delta p_{{\mathrm{min}}}$ is an `equilibrium' value as argued above,
the chaotic contraction is no longer shrinking the scale of phase-space probability 
exponentially and indefinitely as in the non-dissipative setting ( recall $\delta p(t) \sim \delta p(0)e^{-\lambda t}$).
Hence, in principle there would be no Zaslavsky's time $t^*$, but rather, the quantum dissipative evolution
may be well described by the Fokker-Planck type of equation~(\ref{DissWigEq}) at all times, provided that
the initial condition is smooth enough. Importantly, the semiclassical limit is not required for this approximation
to work. 

\section{Contraction vs. diffusion in stochastic dynamics} 
\label{ContrDiff}

Equation~\refeq{DissWigEq} and the discussion from the previous section suggest that,
under suitable conditions, the problem of the dynamics of a quantum system 
connected to an environment may be cast into
the classical evolution of a density according to a Fokker-Planck equation. 
As a consequence, studying the interplay of stretching/contracting dynamics with weak 
noise may also help shed light on quantum dissipation. Particularly interesting scenarios
arise when the deterministic dynamics exhibits chaotic behavior.
It is in fact well known that the phase space of a chaotic system has a self-similar (fractal) structure
 of infinite resolution. However,
in reality, every system experiences noise, coming from experimental uncertainties, neglected
degrees of freedom, or roundoff errors, for example. No matter how weak, noise smoothens out fractals,
giving the system a finite resolution. The consequences are dramatic for the computation of long-time
dynamical averages, such as diffusion coefficients or escape rates, since infinite-dimensional operators
describing the evolution of the system (such as Fokker-Planck) effectively become finite matrices. 
With the aim of efficiently estimating
long-time averages of observables for a chaotic dynamical system affected by background noise, 
a recent endeavor carried on over the past decade has achieved a technique  
to partition the chaotic phase space up to its optimal resolution,
using unstable periodic orbits. 
The benchmark models already treated range from
one-dimensional discrete-time repellers~\cite{LipCvi10}, and general unimodal
maps~\cite{CviLip12}, to two-dimensional chaotic attractors~\cite{HenLipCvi15,noltino}.
Most importantly, a finite resolution for the state space of these models has effectively 
changed the dimensionality of the Fokker-Planck operator from infinite to inherently finite.
Consequently, computations of the desired long-time averages become simpler
and more efficient. 
On a more intuitive note, the present results also bear 
physical significance because,
 even when the external noise is uncorrelated, additive, isotropic, and
homogenous, the interplay of noise and nonlinear
dynamics always results in a local stochastic neighborhood,
whose covariance depends on both the past and the future
noise integrated and nonlinearly convolved with deterministic
evolution along the trajectory. In that sense, noise is effectively never `white' in 
nonlinearity, and thus
the optimal resolution
varies from neighborhood to neighborhood and has to be
computed locally.

 As stated in the introduction, here we attack continuous-time dynamical systems, and  
 begin by studying the evolution of noisy neighborhoods of periodic orbits. 
The simplest yet meaningful models are two-dimensional limit cycles, that can serve as testbed for parsing the
interaction of deterministic dynamics with noise.



\subsection{The Lyapunov equation around a cycle}

\label{LyapEq}
Consider the Fokker-Planck equation

\beq
\partial_t\rho(\xp,t) = -\partial_\xp\left(v(\xp)\rho(\xp,t)\right) +
 \Delta\partial_{\xp\xp}\rho(\xp,t) 
 \,,
\label{DL:FokPl}
\eeq
where $\Delta$ is the diffusion tensor, whose entries are the noise amplitudes 
along each direction ($\Delta$ is diagonal with identical entries for isotropic noise).  
If we look at the dynamics in the neighborhood of a particular deterministic trajectory, we may 
linearize  the velocity field $v(\xp)$ locally, and replace it with $A_a(\xp-\xp_a)$, where $A_a=\frac{\partial v(\xp)}{\partial \xp}\Bigr|_{\xp=\xp_a}$ is the
so-called matrix of variations. Moreover, we may switch to a co-moving reference frame in the desired neighborhood,
say $z_a = \xp - \xp_a$ .

Suppose we start off with an initial density of trajectories of Gaussian shape, that is 
$\msr_{a}(\orbitDist_a) = \frac{1}{C_a}\mathrm{exp}\left(-\transp{\orbitDist}_a {} \frac{1}{\covMat_a}{\,} \orbitDist_a\right)$ .
The short-time solution to~\refeq{DL:FokPl} can then be written in the path-integral form
\bea
\nonumber
\msr_{a+1}(\orbitDist_{a+1})
    &=&
    \frac{1}{C_{a}}\int [d\orbitDist_a] \,
\exp{\left[-\frac{1}{2}\transp{\left(\orbitDist_{a+1}- (\mathbf{1}+A_a\delta t) \orbitDist_a\right)}
         {} \frac{1}{\diffTen\delta t} {\,}
           \left(\orbitDist_{a+1}-(\mathbf{1}+A_a\delta t) \orbitDist_a\right)
                -\frac{1}{2}\transp{\orbitDist}_a {} \frac{1}{\covMat_a}{\,} \orbitDist_a\right]}
    \continue
    &=&
     \frac{1}{C_{a+1}}
        \;
    \exp{\left[-\frac{1}{2}
      {\transp{\orbitDist}_{a+1}} {}
        {} \frac{1}{(\mathbf{1}+A_a\delta t)\covMat_a(\mathbf{1}+A_a\delta t)^T + \diffTen\delta t}   {\,} \orbitDist_{a+1}\right]}
\,.
\label{FP_pathint}
\eea
One can then infer the relation between input and output quadratic forms in the exponential
\beq
\covMat_{a+1} = \Delta\delta t +
(1 + A_{a}\delta t )\covMat_a\transp{(1+A_a\delta t)}
\, ,
\ee{DL:covMat_map}
and, neglecting terms of order $\delta t^2$,  recover the time-dependent Lyapunov equation
\beq
\dot{\covMat} = A(t) \covMat + \transp{\covMat A}(t) + \diffTen
    \,,\qquad
\covMat(t_0) = \covMat_0
\,.
\ee{contLyapODE}


Following the theory of time-dependent ordinary differential equations~\rf{Amann90},
we may write the solution of~(\ref{contLyapODE}) as
\beq
Q(t) = J(t,t_0)\left[Q(t_0) +
\int_{t_0}^t J^{-1}(s,t_0)\diffTen \transp{\left(J^{-1}(s,t_0)\right)}ds\right]\transp{J}(t,t_0)
\,.
\ee{DL:tdLyapSol}
Here $J(t,t_0)$ is the Jacobian along a flow $\xp=\xp(t)$:
\beq
\frac{d}{dt}J(t,t_0) = A(\xp)J(t,t_0),
\hspace{2cm} J(t_0,t_0) = \mathbf{1}
\, .
\ee{DL:j_ev}
One can verify this by just plugging the solution above into the equation.
Alternatively, one can write Eq.~(\ref{DL:tdLyapSol}) in the simpler form 
\beq
Q(t) = J(t,t_0)Q(t_0)\transp{J}(t,t_0) +
\int_{t_0}^t J(t,s)\diffTen \transp{J}(t,s)ds
\,,
\ee{PC:tdLyapSol}
where the notation $J(t,s)$ means that the Jacobian is computed following a trajectory 
that starts at time $s$ and ends at time $t$, consistently with Eq.~(\ref{DL:j_ev}).

\subsection{Noisy circle}
\begin{figure}[tbh!]
\centerline{
\includegraphics[width=6.5cm]{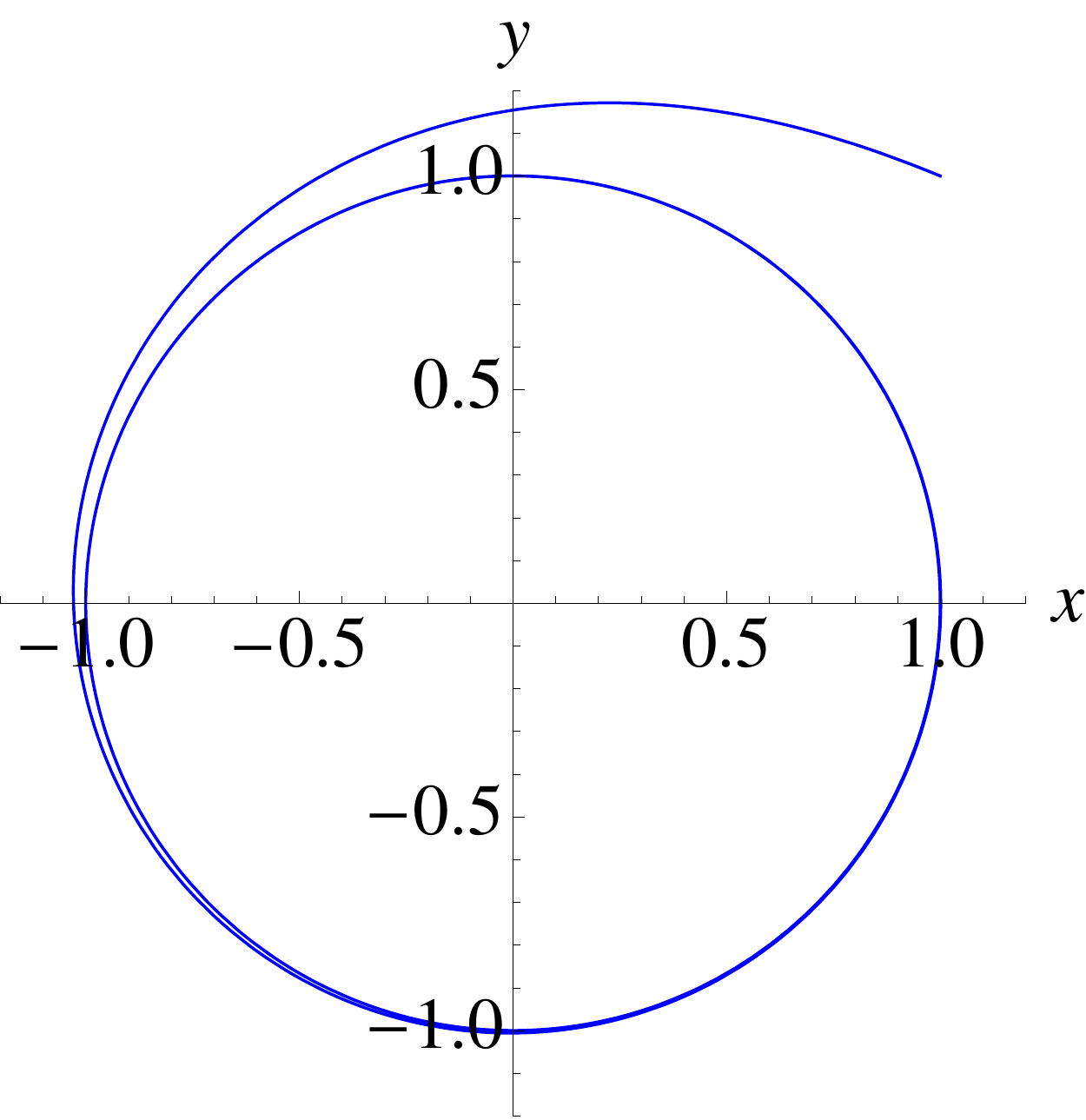}
}
\caption{Solution of the numerically integrated Eq.~(\ref{DL:circle}) without noise. Any initial condition converges to the 
circular limit cycle.}        
\label{limcirc}
\end{figure}    
Next, consider one of the simplest 2-dimensional
dynamical systems, a pair of ODEs with
a circular limit cycle of radius $r_c$,
together with additive isotropic white noise of strength $2D$:
\bea
\dot{x} &=& \Lyap (r_c - \sqrt{x^2 + y^2})x - \omega y + \sqrt{2D}\xi_x
\continue
\dot{y} &=& \Lyap (r_c - \sqrt{x^2 + y^2})y + \omega x + \sqrt{2D}\xi_y
\label{DL:circle}
\eea
where
\beq
<\xi_x(t)\xi_x(\tau)> = \delta(t-\tau), \hspace{5mm}  <\xi_x(t)\xi_y(\tau)> = 0 .
\ee{DL:noise}
In polar coordinates, this is written
\bea
\dot{r} &=& \Lyap(r_c - r)r + \sqrt{2D}\xi_x \cos\theta + \sqrt{2D}\xi_y \sin\theta
        \continue
 \dot{\theta} &=& \omega - \sqrt{2D}\xi_x \frac{\sin\theta}{r} + \sqrt{2D}\xi_y \frac{\cos\theta}{r}
\label{DL:pol_circle}
\eea
This Langevin-type equation produces the drift and diffusion coefficients\rf{Risken96}
\bea
D_r &=& \Lyap(r_c - r)r + \frac{2D}{r}
\continue
D_\theta &=& \omega
\continue
D_{rr} &=& 2D
\continue
D_{\theta\theta} &=& \frac{2D}{r^2}
\label{DL:coefficients}
\eea
which then determine the {\Fokker} equation for the system:
\beq
\partial_t P + \frac{1}{r}\partial_r[\Lyap(r_c- r)rP] + \partial_\theta \omega P - \frac{D}{r}\partial_r(r\partial_r P) -
\frac{D}{r^2}\partial_{\theta\theta}P = 0
\ee{DL:2dFP}
The limit cycle $r=r_c$ can be either stable or unstable
depending on the sign of $\Lyap$. Let us consider the
stable case. 

The first thing to look for is a stationary
solution to the asymptotic form of \refeq{DL:2dFP}:
\beq
\partial_r[\Lyap(r_c-r)rP_\infty]  - D\partial_r(r\partial_r P_\infty) = 0
\ee{DL:asympt}
A solution is
\beq
P_\infty(r) = Ce^{-\frac{\Lyap}{2D}(r-r_c)^2}
\,,
\ee{DL:statsol}
which implies that $P_\infty$ is a Gaussian of width $2\sqrt{D/\Lyap}$  in the neighborhood of the limit cycle.
The general solution to \refeq{DL:2dFP} is\rf{Risken96}
\beq
P(r,\theta,t) = e^{-\frac{\Lyap}{2D}(r-r_c)^2}\sum_{n=0}^\infty\sum_{\nu=-\infty}^\infty
          A_n^\nu e^{-\eigenvL_n^\nu t}(r-r_c)^{|\nu|}L_n^{|\nu|}(r-r_c)e^{i\nu\theta}
\ee{DL:ContTimeSol}
where $L_n^{|\nu|}(r-r_c)$ are generalized Laguerre polynomials and both the eigenvalues $\eigenvL_n^\nu$ and the coefficients $A_n^\nu$ can be found numerically.

\subsubsection{Neighborhood and coordinates}

This problem has an obvious symmetry, which allows us to guess the right (nonlinear!) change of coordinates, as well as the stationary solution, that is 
independent of the angular coordinate. The result is that the noisy neighborhood of the limit cycle is determined by the 
variance of the stationary solution~\refeq{DL:statsol}. In general, however, we might not be so lucky, and guessing a suitable, possibly nonlinear, change
of coordinates is probably beyond our reach. One way to identify a neighborhood for a periodic or any other orbit is to integrate the time-dependent Lyapunov
equation~\refeq{contLyapODE} in the original (Cartesian) coordinates, but in a co-moving frame defined by the local coordinates $z_a=\xp-\xp_a$ introduced in
section~\ref{LyapEq}. 

\begin{figure}[tbh!]
\centerline{
(a)
\includegraphics[width=6.cm]{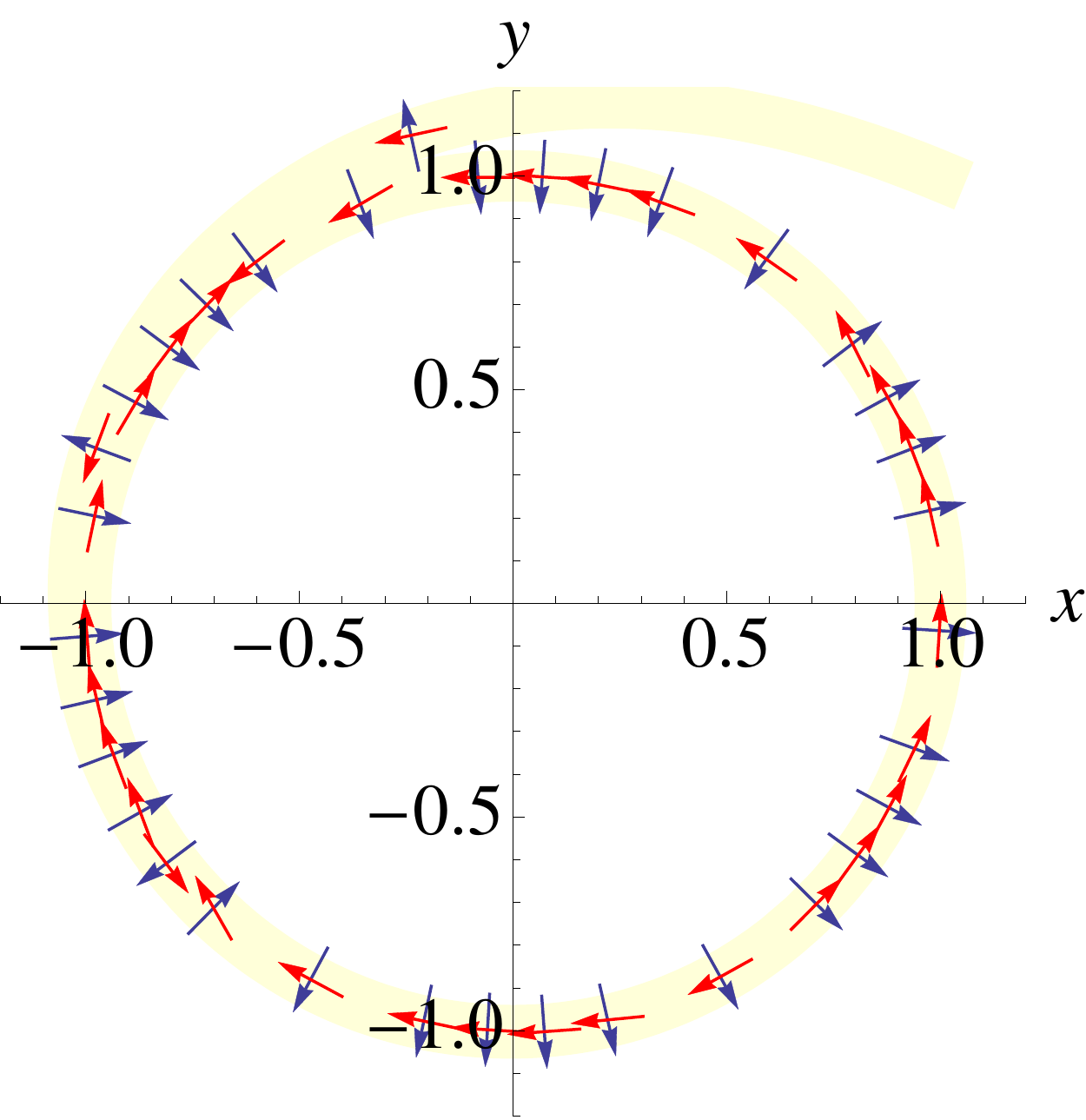}
(b)
\includegraphics[width=6.cm]{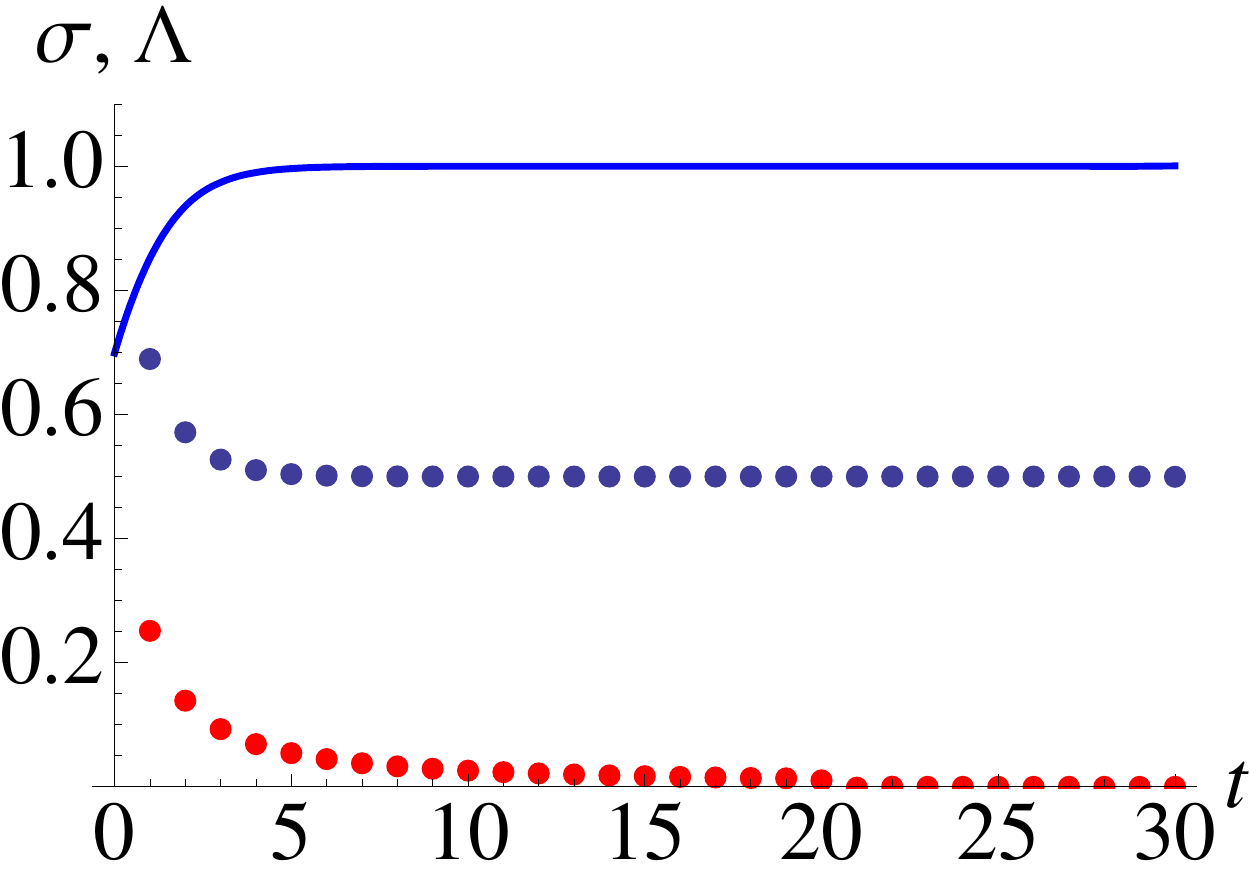}
}  
\vskip 0.2cm
\centerline{
(c)
\includegraphics[width=6.cm]{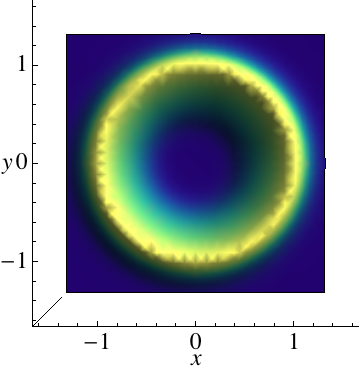}
(d)
\includegraphics[width=6.cm]{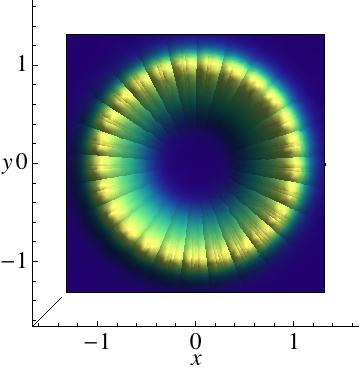}
}
\caption{(a) Solution of the numerically integrated Eq.~(\ref{DL:circle}), together with the eigenvectors (arrows) of the covariant matrix $Q^{-1}$, as given by the
solution~(\ref{PC:tdLyapSol}) of the forward Lyapunov equation, for noise amplitude $2D=0.1$. The light yellow stripe is a pictorial representation of the Gaussian tube along the limit cycle (plan view); (b) The eigenvalues $\Lambda_1$ (blue dots), $\Lambda_2$ (red dots) of $Q^{-1}$, and
the width $\sigma$  (solid line) of the evolved density versus time $t$; (c) The exact steady-state solution~(\ref{DL:statsol}) 
to the Fokker-Planck equation for a noisy circular limit cycle; (d) The approximation to the same steady-state, obtained by piecing together solutions~(\ref{PC:tdLyapSol}) 
to the Lyapunov equation around the limit cycle.}        
\label{giotto}
\end{figure}

Figure~\ref{giotto} illustrates this second approach: the forward Lyapunov equation~(\ref{contLyapODE}) is numerically solved along an orbit that converges to 
the circular limit cycle~\footnote{Here we take the diffusion tensor 
$\Delta= \left(\begin{array}{cc} 2D & 0 \\ 0 & 2D \end{array} \right)$}, 
and its solution~(\ref{PC:tdLyapSol}) is sampled along the trajectory and \textit{inverted} to obtain $Q^{-1}(t)$, the covariance matrix of the 
Gaussian density, that produces a tube (in the figure in light yellow)  
along the orbit. The eigenvalues of $Q^{-1}(t)$ are found to converge to $\Lambda_1=\frac{\lambda}{2D}$, consistently with the result for the 
width of the stationary-state solution~(\ref{DL:statsol}) of the full-blown radial Fokker-Planck equation~(\ref{DL:asympt}): that determines the width of the tube, 
$\sigma=\sqrt{1/2\Lambda_1}$. The second eigenvalue of $Q^{-1}$ is $\Lambda_2=0$, as it appears from
fig.~\ref{giotto}(b). The latter eigenvalue is to be read as follows: while the forward Lyapunov equation converges to a finite limit in the stable (radial) direction,
where noise balances contracting dynamics, it diverges along the marginal (tangent) direction, and therefore its inverse converges to zero, asymptotically.  

As shown in Fig.~\ref{giotto}(c)-(d), the exact solution~(\ref{DL:statsol})  of the 
Fokker-Planck equation is well reproduced by piecing together Gaussian tubes of covariance $Q^{-1}(t)$, each
computed around a definite point of the noiseless limit cycle (the spurious lines orthogonal to the circle in Fig.~\ref{giotto}(d) are due to the finite sampling of the Gaussian tubes, that 
should ideally be a continuum).           

As we may mostly be interested in the solution of the Lyapunov equation near unstable periodic orbits (like in a chaotic system), we then need to solve the
same problem backwards in time, otherwise said by studying adjoint evolution, or the adjoint Lyapunov equation.

\subsection{Adjoint Lyapunov equation}
The backward evolution is described by the adjoint Fokker-Planck equation
 \beq
\partial_t\rho(\xp,t) = v(\xp)\partial_\xp\rho(\xp,t) +
 \Delta\partial_{\xp\xp}\rho(\xp,t) 
 \,.
\ee{DL:adjFokPl}    
Following the line of thought of section~\ref{LyapEq}, we can write the path-integral evolution of a Gaussian density in the neighborhood of an orbit   
\bea
\nonumber
\msr_{a}(\orbitDist_{a})
    &=&
    \frac{1}{C_{a+1}}\int [d\orbitDist_{a+1}] \,
\exp{[-\frac{1}{2}\transp{(\orbitDist_{a+1}- (\mathbf{1}+A_a\delta t) \orbitDist_a)}
         {} \frac{1}{\diffTen\delta t} {\,}
           (\orbitDist_{a+1}-(\mathbf{1}+A_a\delta t) \orbitDist_a)
                -\frac{1}{2}\transp{\orbitDist}_{a+1} {} \frac{1}{\covMat_{a+1}}{\,} \orbitDist_{a+1}]}
    \continue
    &=&
     \frac{1}{C_{a}}
        \;
    \exp{\left[-\frac{1}{2}
      {\transp{\orbitDist}_{a}} {}
        {} \frac{1}{Q_{a}}{\orbitDist}_{a}\right]}
\,,
\label{adjFP_pathint}
\eea
where
\beq
\covMat_a = (\mathbf{1}+A_a\delta t)^{-1}\left(\covMat_{a+1}+\diffTen\delta t\right)\left[\transp{(\mathbf{1}+A_a\delta t)}\right]^{-1}
\,.
\ee{adjQeq} 
Analogously to the forward evolution, we can take the limit of infinitesimal time intervals and get the differential equation
\beq
\dot{Q} = \Delta - A(t)Q - Q\transp{A}(t)
\,,
\ee{adjLyapEq}
the adjoint (or backward-) Lyapunov equation. Compared to the forward Lyapunov equation~(\ref{contLyapODE}),  the adjoint evolution~(\ref{adjLyapEq})
features the `time reversal' operation $A(t)\rightarrow -A(t)$, and therefore we can still use Eq.~(\ref{PC:tdLyapSol}) as a solution, as long as the Jacobian
along the orbit is computed as 
 \beq
\frac{d}{dt}J(t,t_0) = -A(\xp)J(t,t_0),
\hspace{2cm} J(t_0,t_0) = \mathbf{1}
\,,
\ee{DL:backj_ev}
and its computation follows the time reversed flow, that is the solution to the dynamical system $\dot{\xp}=-v(\xp)$ .

\section{Non-circular limit cycles: Classical noise vs. quantum dissipation}
\label{VdP}

We now turn our attention to non-circular limit cycles with background noise,
and determine the steady-state density distribution yielded by the Fokker-Planck equation. We do so by integrating
the Lyapunov equation in the neighborhood of a trajectory that eventually converges to the limit cycle.   

The paradigmatic models of our choice both come from the nonlinear oscillator 
\beq
\ddot{x} + \omega_0^2 x + \mu g( x, \dot{x}) = 0
\,,
\label{DL:nlosc}
\eeq
with $g$ a nonlinear function of position and velocity. 
Equation~(\ref{DL:nlosc}) may be reduced to a dynamical system as the Van der Pol oscillator,
by taking $g=(x^2-b)\dot{x}$,  or as the Rayleigh model, by taking $g=\frac{\dot{x}^3}{3} -b\dot{x}$ in Eq.~(\ref{DL:nlosc}).
In what follows, we shall set $b=3$, $\omega_0=1$, and we will tweak $\mu$.
The Van der Pol oscillator takes the form    
\bea
\nonumber
\dot{x} &=& y \\
\dot{y} &=& -\mu\left(x^2 -3\right)y -x
\,,
\label{VdP2}
\eea
 while the Rayleigh model reads (it is customary to swap $x$ and $y$)
 \begin{samepage}
\bea
\nonumber
\dot{x} &=& y  - \mu\left( \frac{1}{3}x^3 -3x \right) \\
\dot{y} &=& -x
\,.
\label{Rayleigh}
\eea   
\end{samepage}
\begin{figure}[tbh!]
\centerline{
\includegraphics[width=5.5cm]{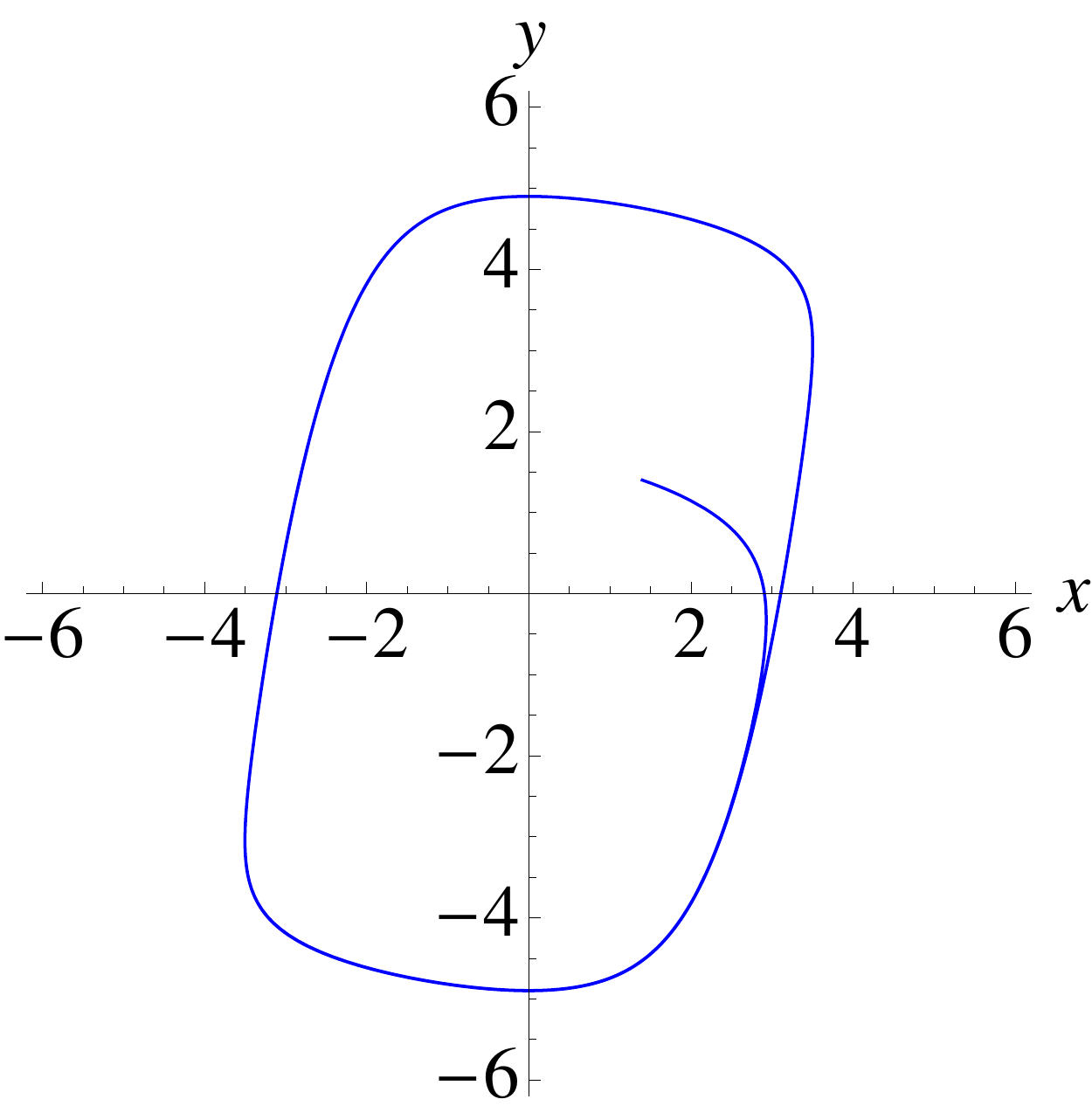}
\hskip 0.2cm
\includegraphics[width=5.5cm]{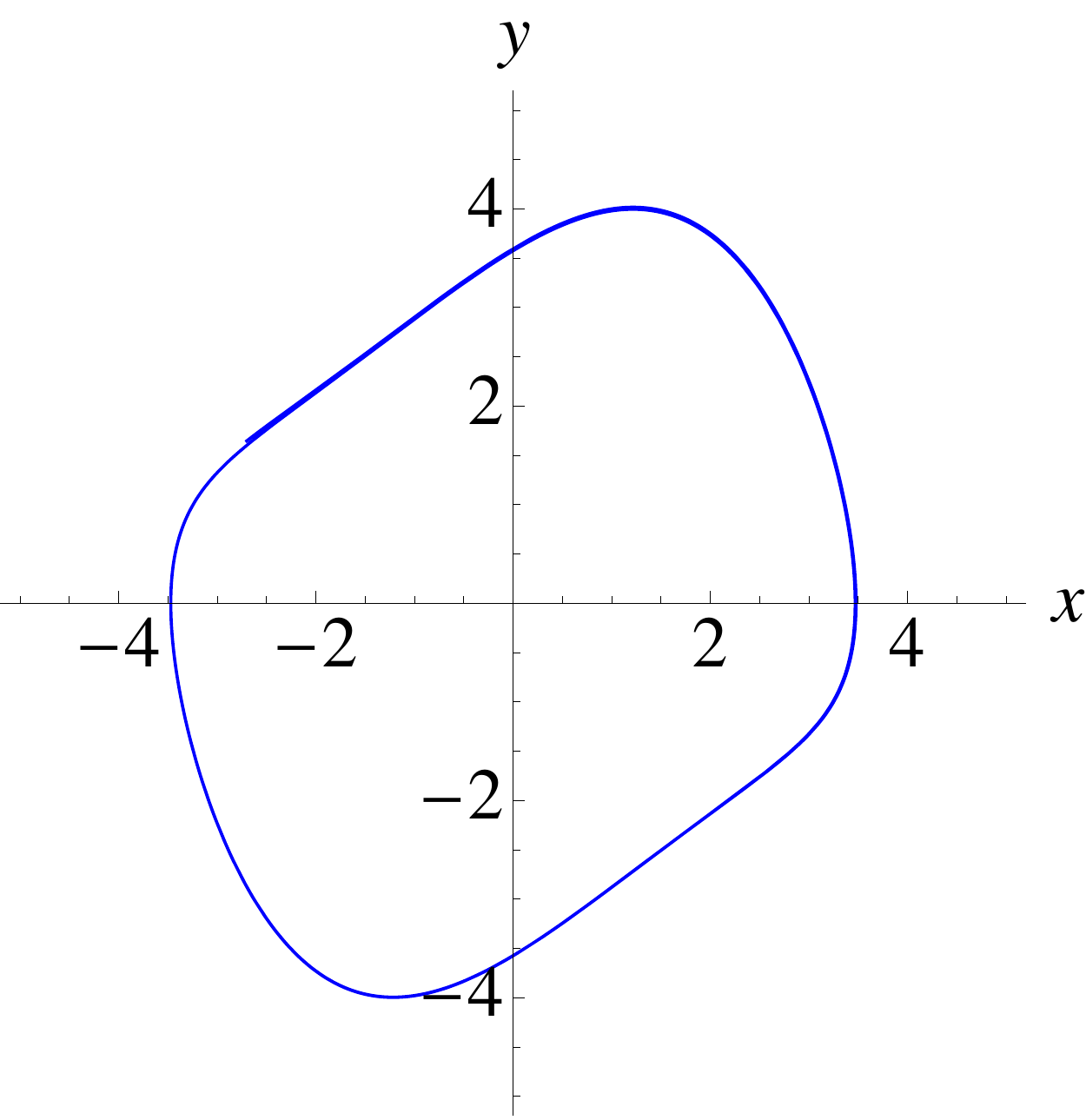}
}
\caption{Solution of the numerically integrated (a) Eq.~(\ref{Rayleigh}), and (b) Eq.~(\ref{VdP2}), without noise. Any initial condition converges to a 
limit cycle.}        
\label{RayleighVdP}
\end{figure}    

\begin{figure}[tbh!]
\centerline{
(a)
\includegraphics[width=7cm]{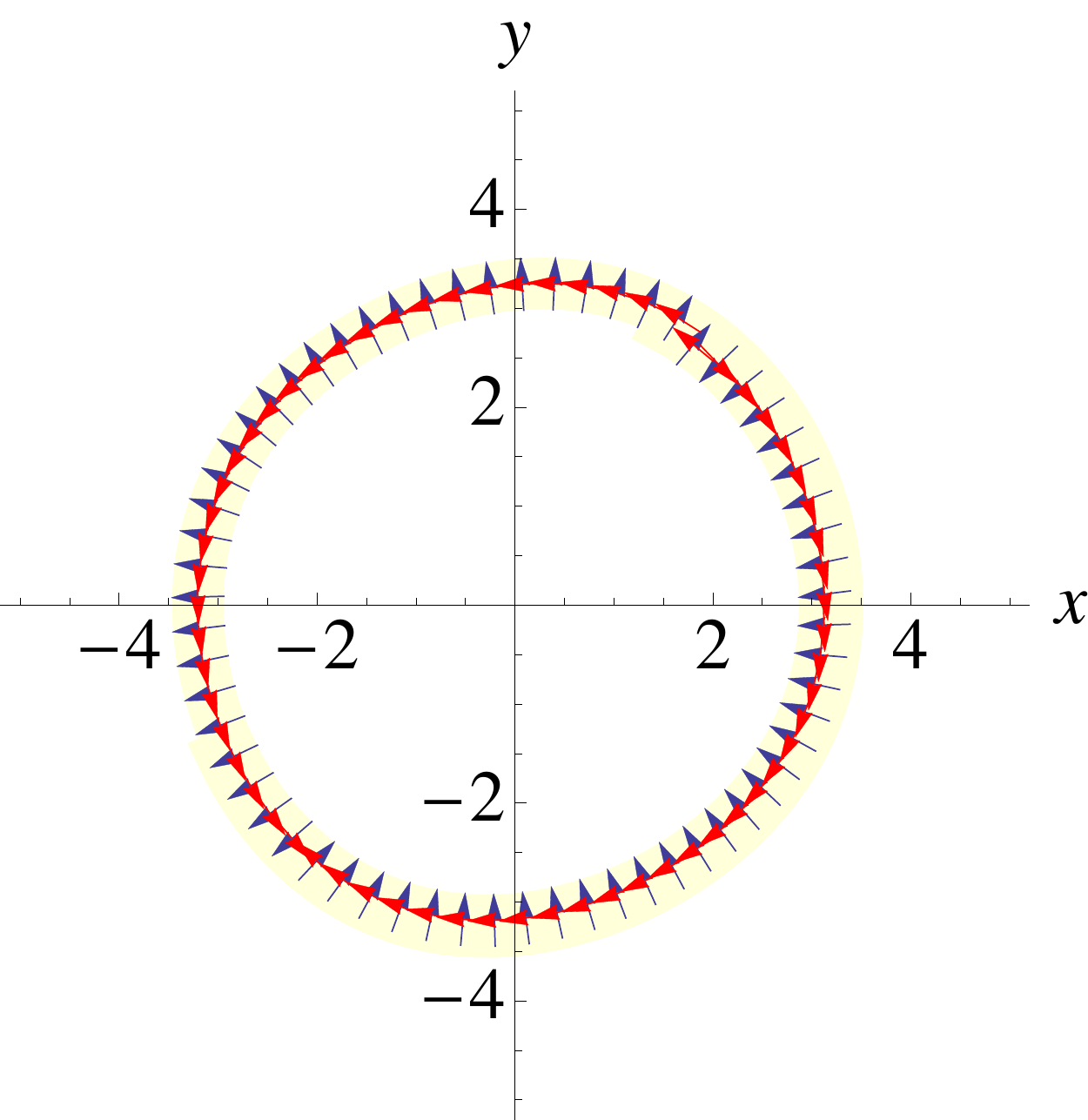}
(b)
\includegraphics[width=7cm]{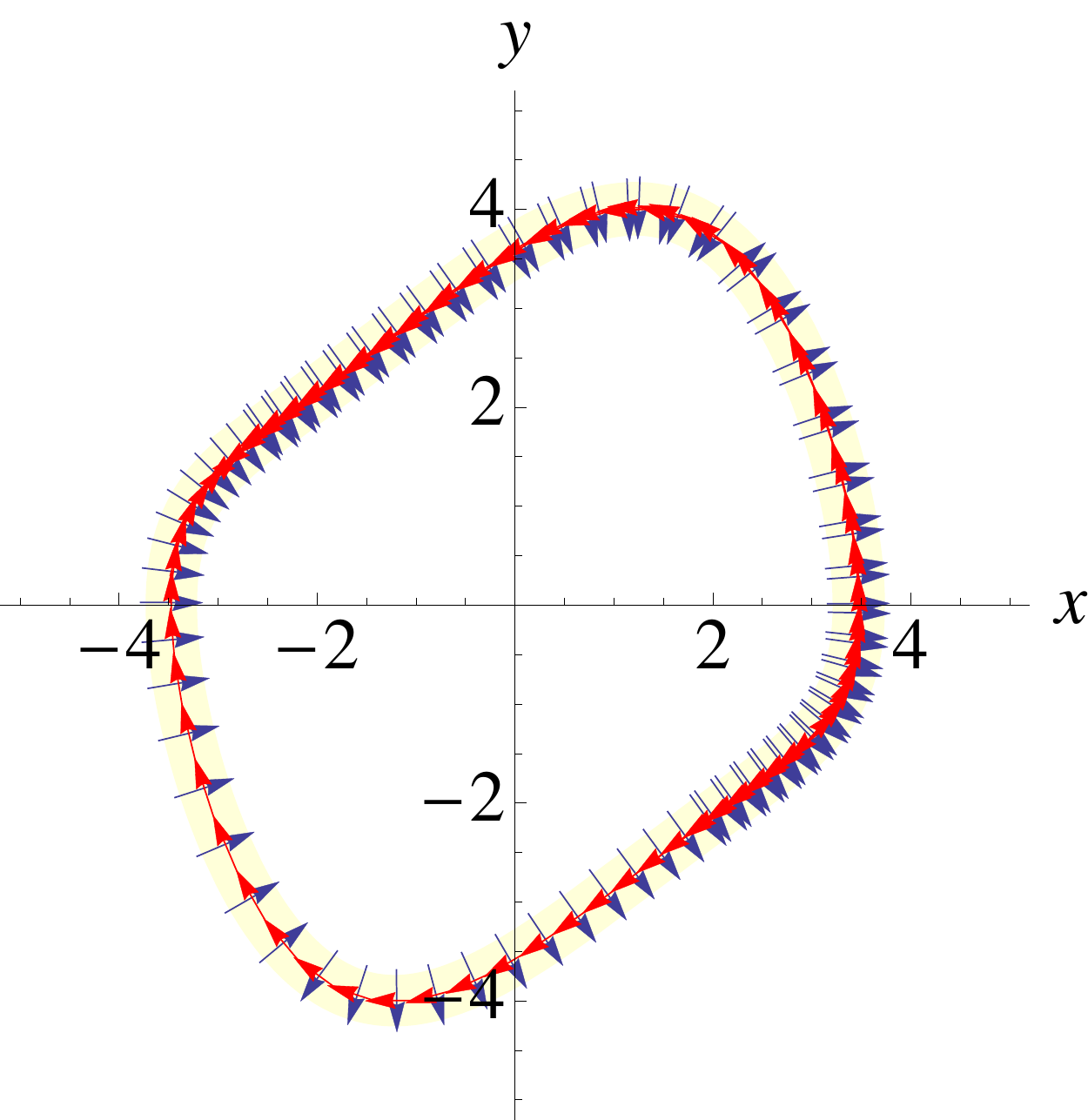}
}
\vskip 0.2cm
\centerline{
(c)
\includegraphics[width=7cm]{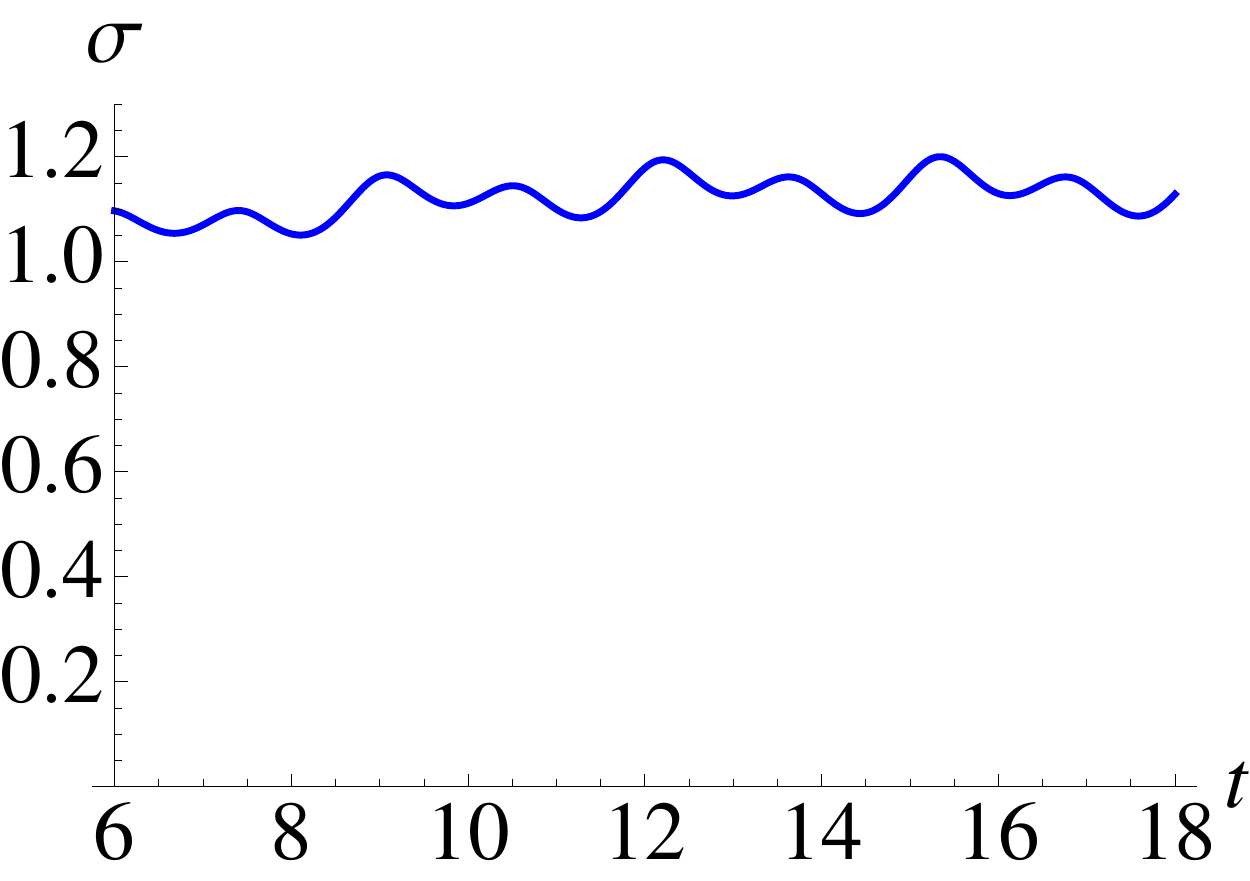}
(d)
\includegraphics[width=7cm]{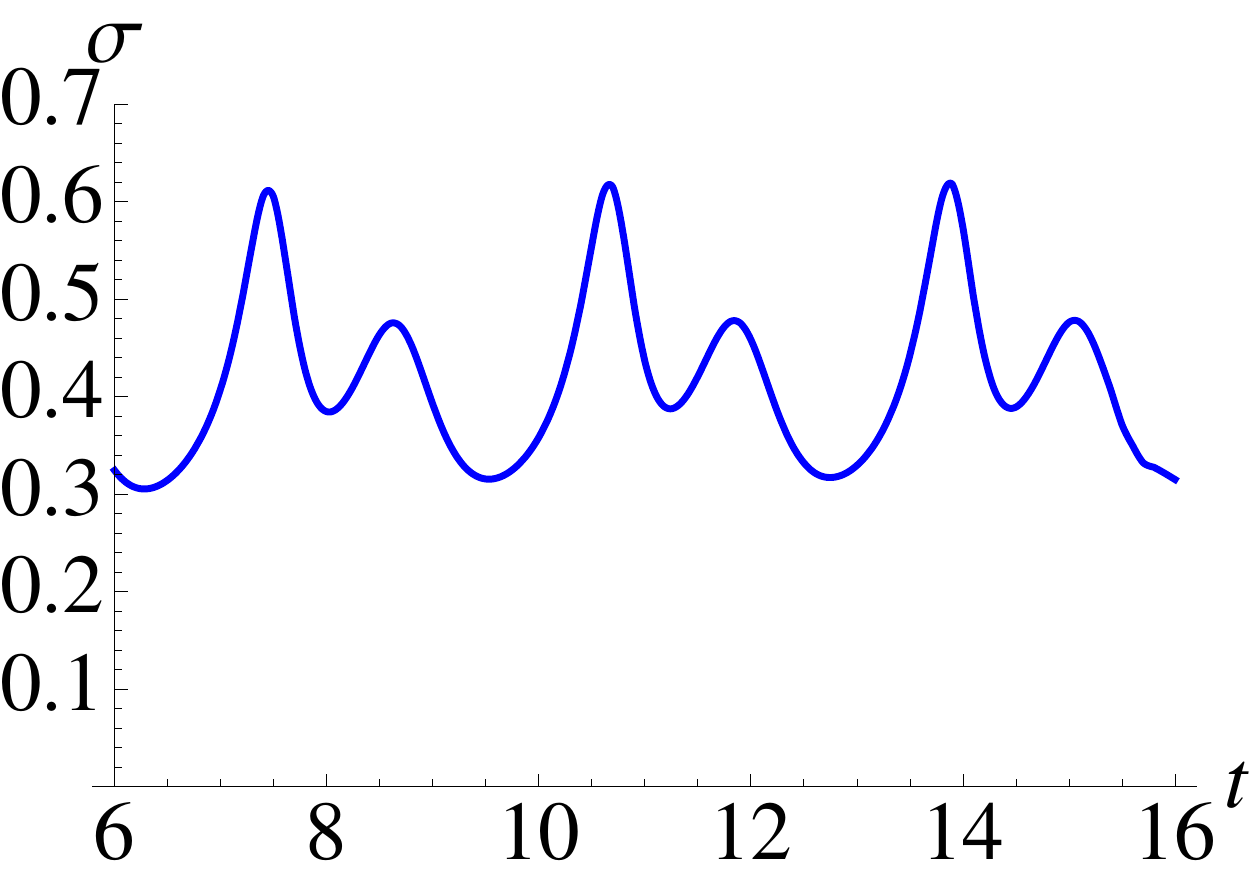}
}
\vskip 0.2cm
\centerline{
(e)
\includegraphics[width=7cm]{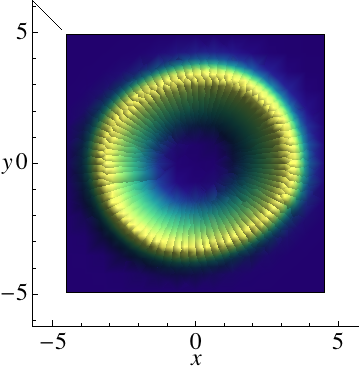}
(f)
\includegraphics[width=7cm]{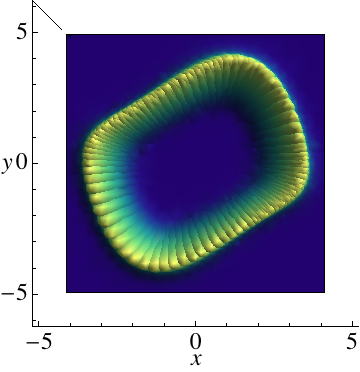}
}
\caption{Top: solution of the numerically integrated Eq.~(\ref{VdP2}), together with the eigenvectors (arrows) of the covariant matrix $Q^{-1}$, as given by the
solution~(\ref{PC:tdLyapSol}) of the forward Lyapunov equation. In Eq.~(\ref{VdP2}), we take (a)-(c)-(e) $\mu=0.03$, and (b)-(d)-(f) $\mu=0.2$,  
while the amplitude of the noise is set to $2D=0.1$.
The light yellow stripe represents the width $\sigma$ of the Gaussian density around the limit cycle; Middle: width of the Lyapunov tube, determined by the nonzero
eigenvalue of $Q^{-1}(t)$, vs. time $t$.  The cycle period is $t_p\approx7$ time units; Bottom: Gaussian solutions of the linearized Fokker-Planck equation along the limit cycle.} 
\label{VdPfig}
\end{figure}

\begin{figure}[tbh!]
\centerline{
(a)
\includegraphics[width=7cm]{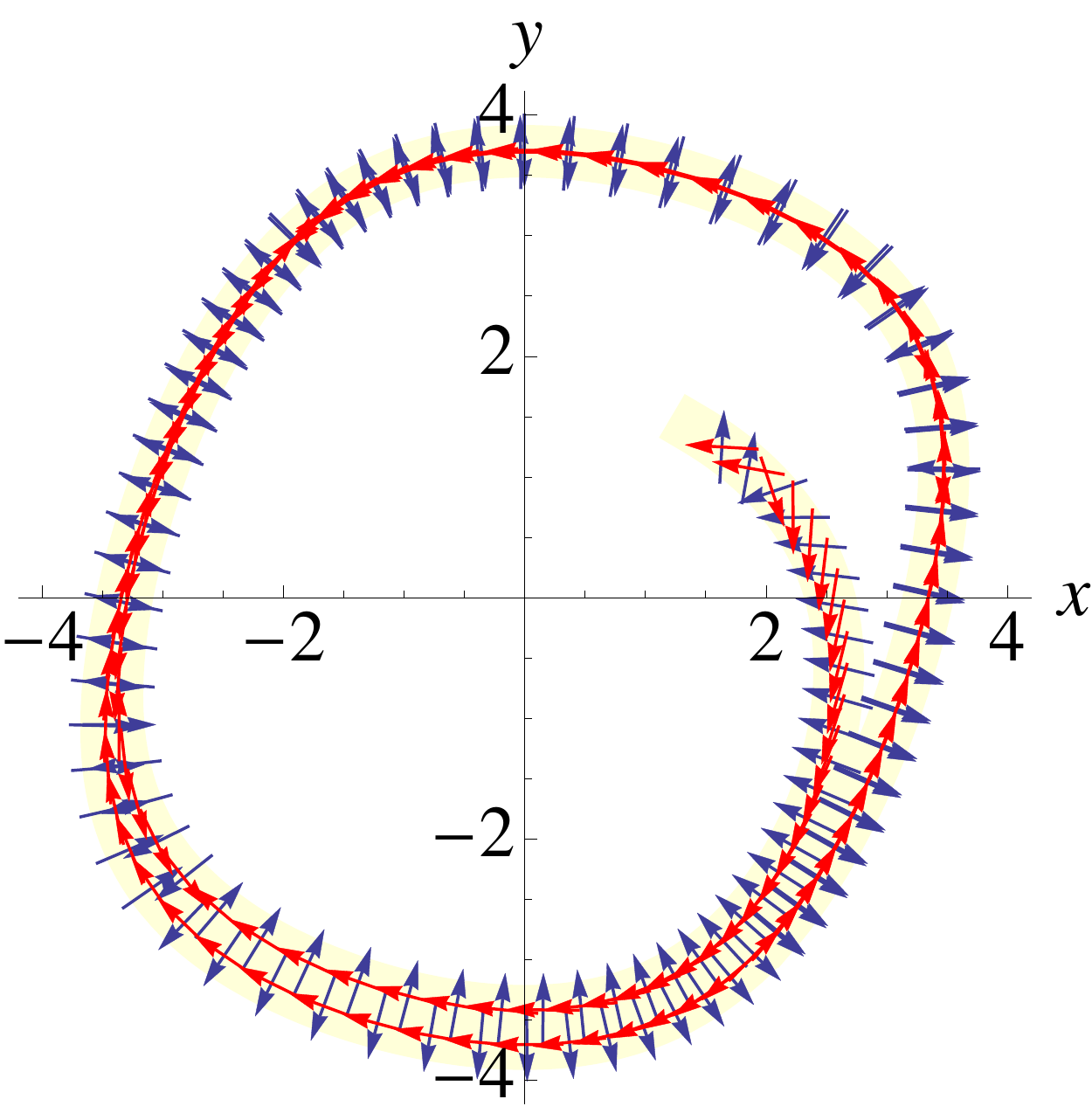}
(b)
\includegraphics[width=7cm]{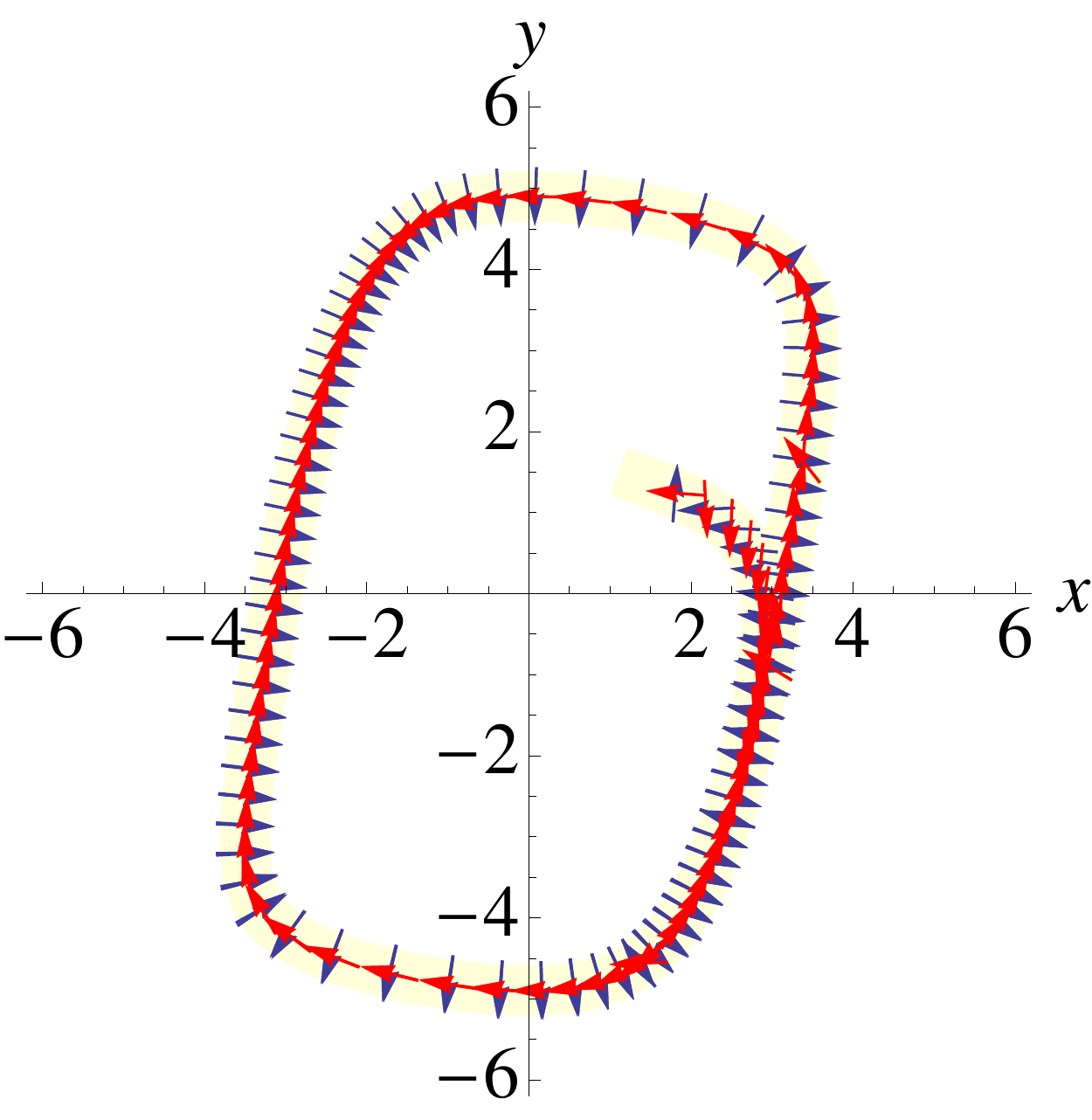}
}
\vskip 0.2cm
\centerline{
(c)
\includegraphics[width=7cm]{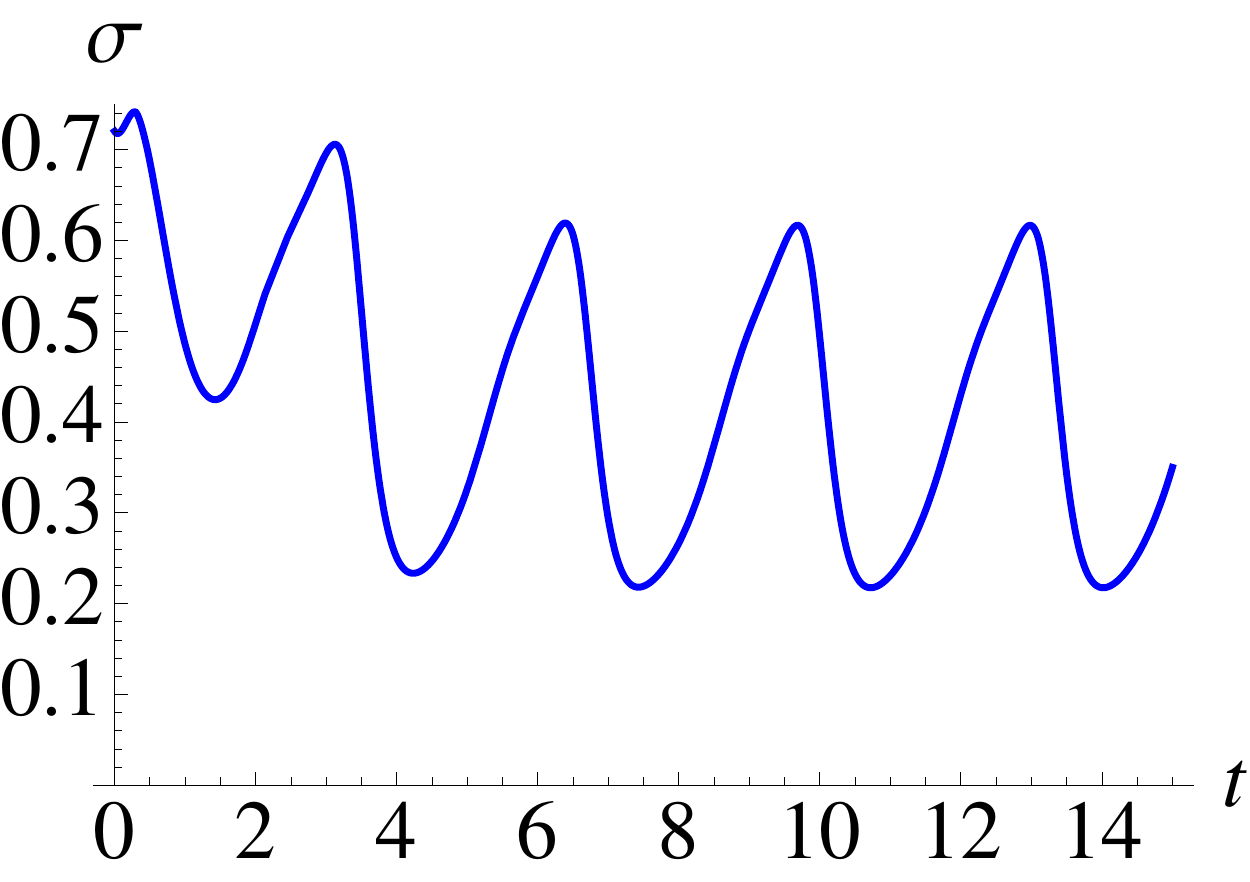}
(d)
\includegraphics[width=7cm]{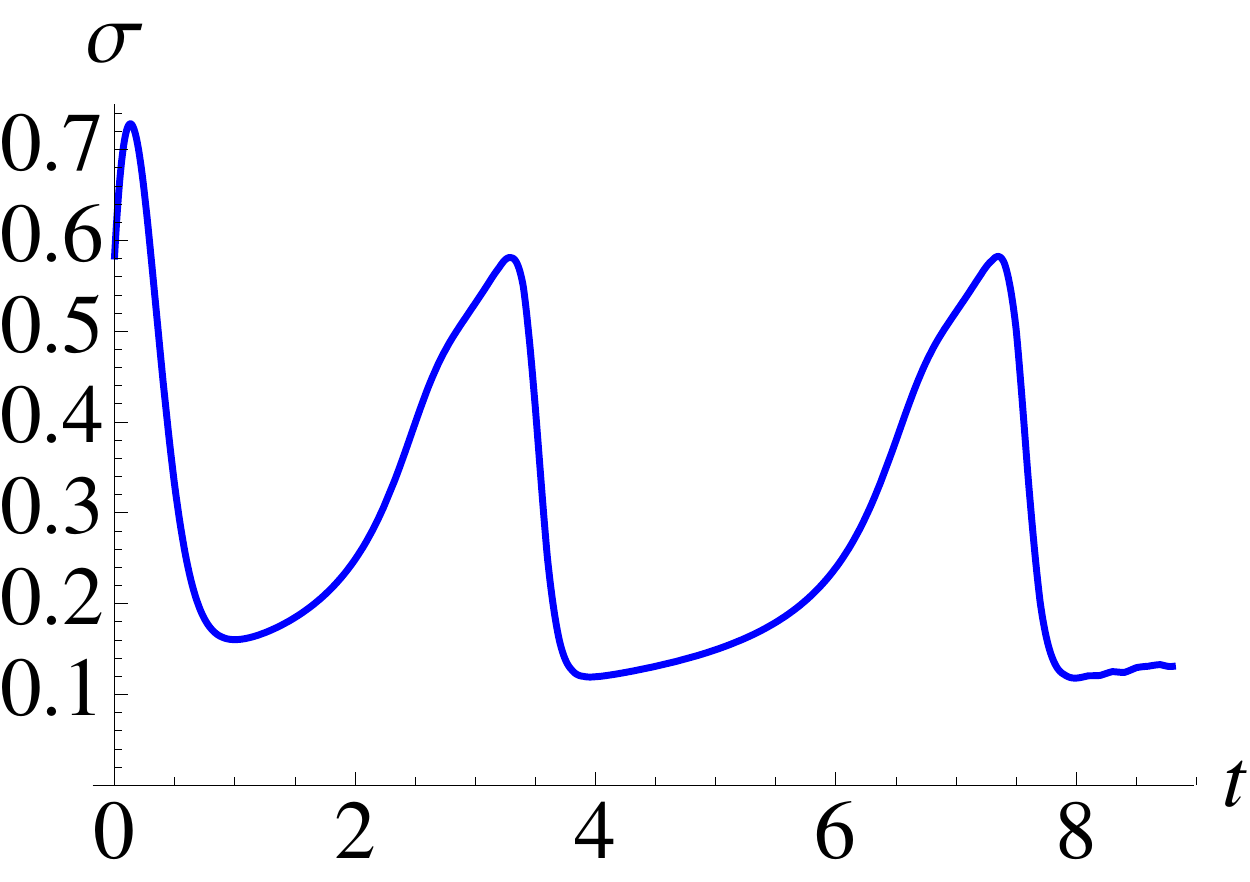}
}
\vskip 0.2cm
\centerline{
(e)
\includegraphics[width=7cm]{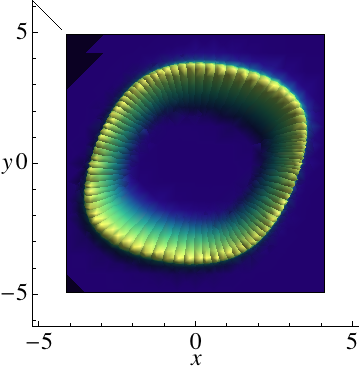}
(f)
\includegraphics[width=7cm]{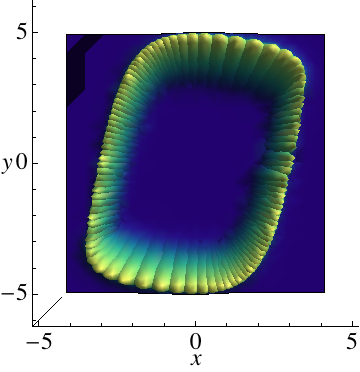}
}
\caption{Top: solution of the numerically integrated Eq.~(\ref{Rayleigh}), together with the eigenvectors (arrows) of the covariant matrix $Q^{-1}$, as given by the
solution~(\ref{PC:tdLyapSol}) of the forward Lyapunov equation. In Eq.~(\ref{Rayleigh}), we take (a)-(c)-(e) $\mu=0.3$, and (b)-(d)-(f) $\mu=0.8$,  
while the amplitude of the noise is set to $2D=0.1$.
The light yellow stripe represents the width $\sigma$ of the Gaussian density around the limit cycle;
 Middle: width of the Lyapunov tube, determined by the nonzero
eigenvalue of $Q^{-1}(t)$, vs. time $t$. The cycle period is $t_p\approx7$ time units; Bottom: Gaussian solutions of the linearized Fokker-Planck equation along the limit cycle.} 
\label{Rayfig}
\end{figure}

In both classical systems, the dynamics converges to a limit-cycle of
non-circular shape, which depends on the parameters, and it is characterized by
fast and slow motions. One therefore expects the interplay of noise with the inherent
nonlinear contraction to be non-uniform, unlike in the circular limit cycle examined in
the previous section, and to give rise to a stationary density distribution of varying   
covariance along the cycle. We consider both models~(\ref{VdP2}) and~(\ref{Rayleigh}) for 
different values of the parameter $\mu$, so as to gradually increase the eccentricity of 
the limit cycle, from a deformed circle [fig.~\ref{VdPfig}(a)] to a nearly rectangular orbit [fig.~\ref{Rayfig}(b)],
where the deterministic stretching/contraction are most inhomogeneous along the cycle.
The most notable feature 
of the covariance of the Gaussian solution 
of the linearized Fokker-Planck equation, denoted by $\sigma$ in figs.~\ref{VdPfig}(c)-(d) and~\ref{Rayfig}(c)-(d),
is its oscillation along the direction orthogonal to that of noiseless motion.  
The more eccentric the limit cycle, the more widely and rapidly $\sigma$ oscillates. That translates to a
Gaussian steady-state density featuring a width that increasingly depends on the position along the orbit with
the parameter $\mu$, as we can see in the three-dimensional/density plots of figs.~\ref{VdPfig}(e)-(f), and~\ref{Rayfig}(e)-(f).
As anticipated, these monodromic Gaussian distributions computed by means of the Lyapunov equation and
portrayed in the figures would be, in a chaotic setting, the building blocks of a partition of the noisy phase space, whose non-uniform resolution
is determined by their overlaps.               

\begin{figure}[!htb]
\centerline{
\includegraphics[width=7.5cm]{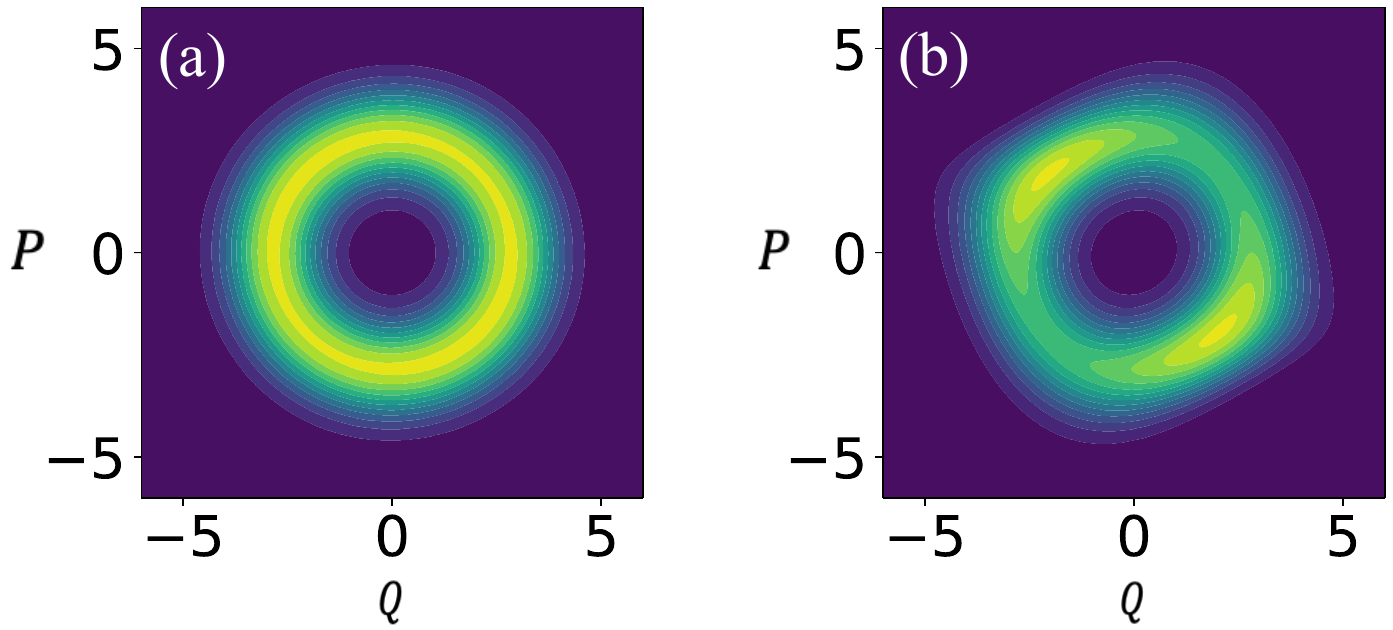}
\hskip 0.3cm
\includegraphics[width=7.5cm]{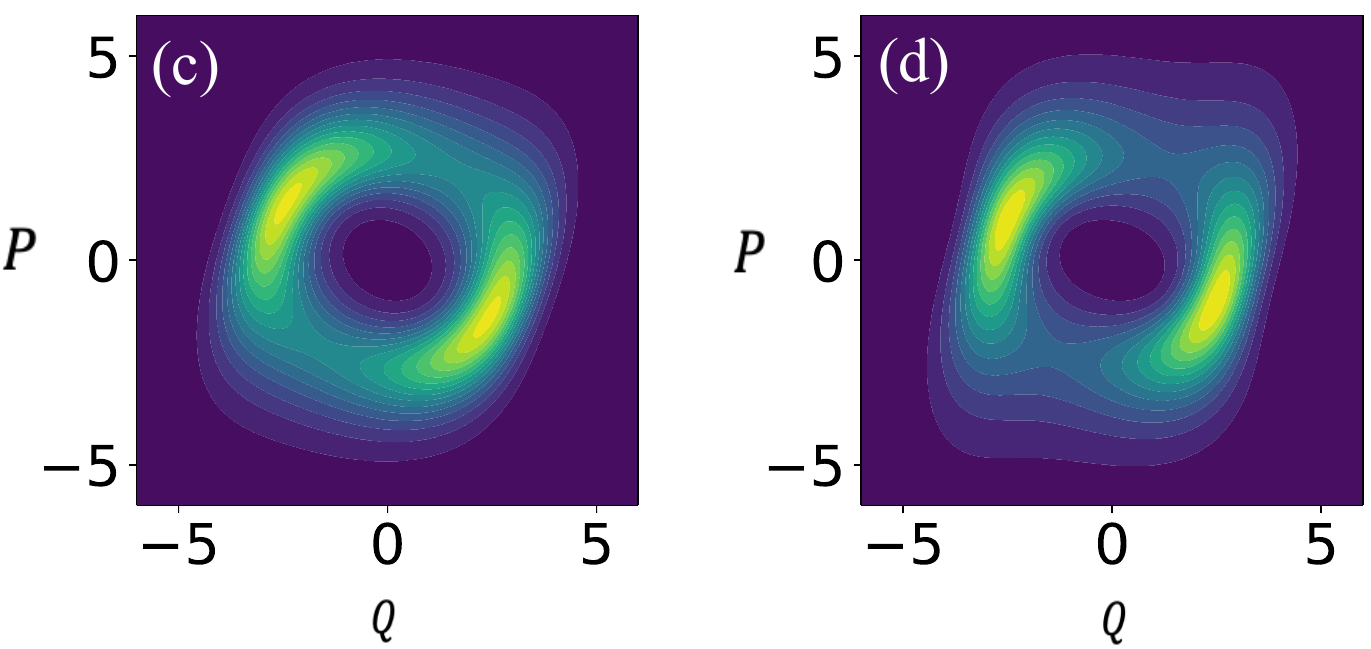}
}  
\caption{Quantum  [(a)-(b)] Van der Pol- and [(c)-(d)] Rayleigh oscillators from Reference~\cite{QVdP20}. The steady-state Wigner function (density plot)  localizes around
limit cycles of increasing eccentricity
[controlled by the parameter $\mu$ in the classical models~(\ref{VdP2})  and~(\ref{Rayleigh})]. Courtesy of A. Chia.}
\label{QVdP}
\end{figure}

The 
Gaussian solutions of the Lyapunov equation computed and illustrated here 
share common traits with the 
steady-state Wigner function of the same two oscillators~(\ref{VdP2}) and~(\ref{Rayleigh}),
as obtained from a fully quantum mechanical computation that has recently appeared in the
literature~\cite{QVdP20}. In that work,  the quantization
is performed by means of creation/annihilation operators
\begin{equation}
\hat{a} = \frac{1}{2}\left(\hat{x}+i\hat{y}\right)
\,,
\end{equation}
and its adjoint $\hat{a}^\dagger$, while dissipative terms  are
added to the Liouville-von Neumann equation~(\ref{LvN}), in the spirit of Lindblad's formalism (in units of $\hbar$):
\begin{linenomath}
\begin{equation}
 \rho_t = -i\left[\hat{H}, \rho\right]  + \sum_j\alpha_j{\cal{D}}\left[f_j(\hat{a}^\dagger, a)\right]\rho 
\,,
\label{LindVdP}
\end{equation}
\end{linenomath}
where 
\begin{linenomath}
\begin{equation}
{\cal{D}}\left[\hat{c}\right]\rho = \hat{c}\rho\hat{c}^\dagger - \frac{1}{2}\hat{c}^\dagger\hat{c}\rho
- \frac{1}{2}\rho\hat{c}^\dagger\hat{c}  
\,,
\label{LindTerm}
\end{equation}
\end{linenomath}  
while the coefficients $\alpha_j$ are functions of the parameters $\omega_0$ and $\mu$, the
Hermitian Hamiltonian $\hat{H}$ has two distinct expressions for the Rayleigh and the Van der Pol oscillators, and, like $f_j$, 
it is a function of linear (e.g. $\hat{a}^\dagger$), bilinear ($\hat{a}^\dagger\hat{a}$) and nonlinear ($\hat{a}^2$) terms involving the creation/annihilation operators
(see ref.~\mbox{
\cite{QVdP20} }
for details). 
The above equation~\refeq{LindVdP} was numerically integrated, and the 
 Wigner function was then found to eventually concentrate around the classical limit cycles, that feature similar eccentricities to the 
 ones considered in the present work and plotted in figs.~\ref{VdPfig} and~\ref{Rayfig}. 
In particular (Fig.~\ref{QVdP}), the steady-state Wigner distribution
  is enhanced along `tubes' of varying width, as it can be noticed in the more eccentric density plots of the Rayleigh model [Fig.~\ref{QVdP}(c)-(d)]. This feature is especially apparent in 
  Fig.~\ref{QVdP}(d), where the high-density region is narrower along the vertical segments of the limit
  cycle (faster classical motion), and wider along its horizontal segments (slower classical motion). 
   It compares directly with the Gaussian solution of the Lyapunov equation portrayed in Fig.~\ref{Rayfig}(f).
    
 It is noted that in the cited work, the authors did not integrate the Wigner equation, but the Lindblad equation with a full-fledged quantum-mechanical algorithm. In particular,    
 the localization of the Wigner function around the classical limit cycles is not to be taken for granted, and it
  legitimates the parallel between their steady-state solutions and the local Gaussian tubes obtained in the present work from the noisy classical system.
  
 On the other hand, the numerical steady-state solutions to Eq.~(\ref{LindTerm}) obtained by the authors of~\cite{QVdP20} are clearly not Gaussians
 centered at the limit cycles (except in Fig.~\ref{QVdP}(a), the case of least eccentricity), as demonstrated by their varying intensities along the periodic orbits. In that sense, the Gaussian  
 Ansatz that turns the Fokker-Planck- into the Lyapunov equation carries but limited information on the phase-space density distributions at equilibrium. Therefore,    
 the analogy proposed here should be taken with a grain of salt, and only considered as a hint for the noisy-classical to
 quantum-dissipative correspondence in a particular system with nontrivial interplay of contraction and diffusion. 
 
Finally, we would like to briefly comment on the difference between classical and quantum dissipation in the present models featuring limit cycles.
 In the classical system the dissipation produces damping, which is then balanced by the noise-induced diffusion. Instead, the quantum dissipation, generated by the
 characteristic Lindblad terms in Eq.~\refeq{LindVdP}, is responsible for both the `friction' that drives densities to localize along
 the classical limit cycles, and the diffusion that spreads out the steady-state Wigner density distribution in the same region of the attracting orbits.
 This is consistent with the more general picture of section~\ref{OpSyst}, where the quantum dissipation brings about both a damping- and a diffusive term
 in the Wigner equation.    

\section{Summary and discussion}
Having reviewed the parallels between the problem of dynamical evolution of a quantum system subject to dissipation and that of a stochastic 
process ruled by Fokker-Planck's equation, we have narrowed our attention down to chaos, and, in particular, to the problem of an inherent 
scale resolution of the phase space. The issue is measuring the conjugated variables  down to a certain precision, which may be set by 
the balance of the contraction rate of the classical chaos with the coupling to the environment, the source of dissipation.

Using the analogy with the problem of classical chaotic dynamics with background noise, we consider the Fokker-Planck equation,
and study its local solutions in the neighborhood of a periodic orbit, that effectively give the latter a finite width, in the phase space.
Solving the problem for two-dimensional limit cycles, as done here, is the starting point: in a chaotic setting, a number of periodic tubes of         
finite width that proliferate exponentially with their length must end up overlapping, and thus determine the finest resolution for the
noisy/quantized state space, that is expected to be non-uniform, as chaos interacts differently with diffusion/dissipation everywhere,
in general. 

The analysis performed in this venue shows that the problem is tractable, and it provides the basic technology to attack it.
Complications and obstacles are ahead for higher-dimensional systems, where stable, unstable, and marginal directions coexist along the 
same orbit, and where the solution to the adjoint Fokker-Planck operator introduced here will almost certainly be instrumental to the method.
Still, the progress already achieved by periodic orbit theory in such complex models as the Kuramoto-Sivashinsky or the Navier-Stokes
equation give us confidence in the feasibility of the optimal partition hypothesis in higher-dimensional chaos.

\end{document}